\input{psfig}
\documentclass[]{aa}
\usepackage{graphicx}
\usepackage{deluxetable}
\usepackage{aalongtable}
\begin{document}
\title{RACE-OC Project:\\ Rotation and variability in the $\epsilon$ Chamaeleontis,  Octans, and Argus stellar associations \thanks{The online Tables \ref{tab_period}-\ref{argus_lit} and on-line  Figs.\,\ref{eps_cha_fig1}-\ref{ic2391_fig6} will be available in electronic form at the CDS via anonymous ftp to cdsarc.u-strasbg.fr (130.79.128.5)
or via http://cdsweb.u-strasbg.fr/cgi-bin/qcat?J/A+A/} \thanks{Based on the All Sky Automated Survey (ASAS) and Wide Angle Search for Planets (SuperWASP) photometric data.}}
\author{S.\,Messina\inst{1}
\and          S.\,Desidera\inst{2}
\and          A.\,C.\,Lanzafame\inst{1,3}
\and          M.\,Turatto\inst{1,4}
\and          E.\,F.\,Guinan\inst{5}
}
\offprints{Sergio Messina}
\institute{INAF-Catania Astrophysical Observatory, via S.Sofia, 78 I-95123 Catania, Italy \\
\email{sergio.messina@oact.inaf.it}
\and   
INAF-Padova Astronomical Observatory, Vicolo dell'Osservatorio 5, I - 35122 Padova, Italy  \\
\email{silvano.desidera@oapd.inaf.it}
\and   
University of Catania, Dept. of Physics and Astronomy, via S.Sofia, 78 I-95123 Catania, Italy\\
\email{alessandro.lanzafame@oact.inaf.it}
\and   
INAF-Trieste Astronomical Observatory, Via Tiepolo, 11 I - 34143 Trieste, Italy  \\
\email{massimo.turatto@oats.inaf.it}
\and   
Dept. of Astronomy and Astrophysics, Villanova University, Villanova, 19085, PA, USA\\
\email{edward.guinan@villanova.edu}
\\}

\date{}
\titlerunning{Rotation and variability in young associations}
\authorrunning{S.\,Messina et al.}
\abstract {Rotational properties of late-type low-mass members of associations of known age provide a fundamental source of information on stellar internal structure and its evolution.}
{We aim at  determining the rotational and magnetic-related activity properties of stars at different stages of  evolution. We focus our attention primarily on members of young stellar associations of known ages. Specifically, we extend our previous analysis in Paper I (Messina et al. 2010, A\&A 520, A15) to 3 additional young stellar associations beyond 100 pc and with ages in the range 6-40 Myr: $\epsilon$ Chamaeleontis ($\sim$6 Myr), Octans ($\sim$20 Myr), and Argus ($\sim$40 Myr). Additional rotational data of $\eta$ Chamaeleontis and IC\,2391 clusters are also considered.} 
{Rotational periods were determined by applying the Lomb-Scargle periodogram technique to
photometric time-series data obtained by the All Sky Automated Survey (ASAS) and the Wide Angle Search for Planets (SuperWASP) archives. The
magnetic activity level was derived from the amplitude of the V
light curves.}{We detected the rotational modulation and measured the rotation periods of 56 stars for the first time, confirmed  11 and revised 3 rotation periods already known from the literature. Adding the periods of 10 additional stars retrieved from the literature we determined a sample of 80 periodic stars at ages of $\sim$6, $\sim$20, and $\sim$40 Myr. Using the SuperWASP data we also revisited some of the targets studied in Paper I.}{With the present study we have completed the analysis of the rotational properties of the late-type members of all known young  loose associations in the solar neighbourhood. Considering also the results of Paper I, we have derived the rotation  periods of 241 targets: 171 \it confirmed, \rm 44 \it likely\rm, 26 \it uncertain\rm. The period of the remaining 50 stars known to be part of loose associations still remains unknown.
The rotation period distributions we provided in the 0.8-1.2  M$_{\odot}$ mass range span nine different ages from 1 to $\sim$ 100 Myr. This rotation period catalogue, and specifically the new information presented in this paper at $\sim$6, 20, and 40 Myr, contributes significantly to a better observational description of the angular momentum evolution 
of young stars. The results of the angular momentum evolution model based on this period database will@{\hspace{.1cm}} be
presented in forthcoming papers.}
\maketitle
\rm

\begin{table*}[h,t]
\caption{Summary of the  associations under study.\label{tab-list}}
\begin{tabular}{l l ccccccc}
\hline
Association & Abbrev. & Age  & Distance & Known   & Late-type & Periodic    &  Literature & New Period \\
            &        & (Myr)& (pc)     & Members & Members   & Members    &  Period     & (+Revised) \\
\hline
 \object{$\epsilon$ / $\eta$  Chamaeleontis}	& $\epsilon$ / $\eta$ Cha	          & 6     & 108 & 41 & 30 &  14(25)     & 9  &   10 (+2)\\
 \object{Octans} 							& Oct 				  & 20? & 141 & 15 & 12 &  8(10)     & 0  &  9 (+0)\\
 \object{Argus / IC\,2391} 					& Arg / IC\,2391				  & 40   & 106 & 65 & 57 & 23(45)  & 1 &  37 (+1)\\
 \hline
\end{tabular}
\end{table*}

\section{Introduction}

Rotation period measurements in open clusters and young associations provide a fundamental source of information on the angular momentum evolution of late-type stars. Stars in a cluster represent the most homogeneous sample (in age, initial chemical composition, and interstellar reddening) that can be empirically identified and are therefore essential for our understanding of stellar evolution. The identifications of young nearby stellar loose associations (e.g., Torres et al. \cite{Torres08}; Zuckerman \& Song \cite{Zuckerman04}, and references therein) have further extended the homogeneous stellar samples on which the early phases of stellar evolution can be investigated effectively. Nearby young associations are valuable targets for the characterisation of the stellar rotation and of the link with other stellar and environment properties. Like open clusters, in fact, physical association allows us to study ensembles of stars with similar age and native environment and reduce the uncertainties due to intrinsic dispersion (still of unknown origin) of rotation period, light element’s abundance, and magnetic activity. Such systems are targets of several dedicated studies like spectral characterisation and searches for planets and circumstellar discs (e.g. Torres et al. \cite{Torres06}; Setiawan et al. \cite{Setiawan08}; Chauvin et al. \cite{Chauvin10}; Carpenter et al. \cite{Carpenter09}).\\
\indent
In recent years, a number of valuable monitoring projects of young and intermediate-age open clusters (see, e.g., Herbst \& Mundt \cite{Herbst05}; Herbst et al. \cite{Herbst07}; Hodgkin et al. \cite{Hodgkin06}; von Braun et al. \cite{vonBraun05}) have provided enough data for statistical analyses of the rotational evolution of solar-like stars. Irwin \& Bouvier (\cite{Irwin09}) report a compilation of 3100 period measurements available then. More recent measures of photometric rotation periods include those of  575 stars in M37 by Hartman et al. (\cite{Hartman09}), 368 Pleiades stars by Hartman et al. (\cite{Hartman10}), 122 stars in M37 by Messina et al. (\cite{Messina08}), 38 stars in M11 by Messina et al. (\cite{Messina10}), 46 stars in Coma Berenicis by Collier Cameron et al. (\cite{CollierCameron09}), and 55 stars in M34 by James et al. (\cite{James10}). Extensive survey work in clusters at 1-2 Myr, mainly the Orion Nebula Cluster (ONC) and NGC 2264, have established the initial distribution of rotation rates in low-mass stars (Stassun et al. \cite{Stassun99}; Herbst et al. \cite{Herbst01}; Rebull \cite{Rebull01}; Makidon et al. \cite{Makidon04}; Lamm et al. \cite{Lamm04}). Herbst \& Mundt (\cite{Herbst05}) estimate that 50-60\% of the stars on convective tracks are released from any locking mechanisms very early on and spin-up as a consequence of their contraction in the pre-main sequence (PMS), and these stars account for the rapidly rotating young main sequence stars. The other 40-50\% lose substantial amounts of angular momentum during the first few million years, and end up as slowly rotating main sequence stars. The duration of the rapid angular momentum loss phase is $\sim$5-6 Myr, which is roughly consistent with the lifetime of discs estimated from infrared surveys of young clusters. This supports the hypothesis that the interaction with a circumstellar disc drains angular momentum from the star, thus delaying its spin up for the (variable) duration of its lifetime (e.g., Haisch et al. \cite{Haisch01}). This process is not understood in detail yet (see, e.g., Collier Cameron \& Campbell \cite{CollierCameron93}; Shu et al. \cite{Shu94}; Matt et al. \cite{Matt10}) and is usually modelled by means of very simplified assumptions, e.g., the disc-locking hypothesis (K\"oenigl \cite{Koenigl91}; Collier Cameron et al. \cite{CollierCameron95}) or accretion-driven stellar wind (e.g., Matt \& Pudritz \cite{Matt95}).\\
\indent
The initial distribution of rotation rates and the strong rotational regulation in the first $\sim$5Myr are deemed responsible for the dispersion of rotation periods at a given mass seen in individual young clusters from the PMS until the age of the Hyades. For the Hyades, a narrow correlation between the B$-$V colours and the rotation periods of F, G, and K stars has been found (Radick et al. \cite{Radick87}).  A small scatter of the individual stellar rotation rates about the mean period-colour relation has also been found on M37 (Messina et al. \cite{Messina08}; Hartman et al. \cite{Hartman09}), M35 (Meibom, Mathieu, \& Stassun \cite{Meibom09}), and Coma Berenicess (Collier Cameron et al. \cite{CollierCameron09}). The convergence of the spin rates in $\sim$600 Myr is a consequence of the strong dependence on stellar rotation rate of the angular momentum loss via a thermally driven magnetically channelled wind (e.g., Mestel \& Spruit \cite{Mestel87}; see also Weber \& Davis \cite{Weber67}; Kawaler \cite{Kawaler88}; Chaboyer et al. 1995a,b). For such a convergence to occur at the age of the Hyades, the timescale for coupling the star’s radiative interior to its outer convective zone must also be significantly shorter than the Hyades age (see, e.g., Collier Cameron et al. \cite{CollierCameron09}).\\
\indent
The relationship between age, rotation, and activity has been a crucial topic of stellar evolution over the past 40 years. Angular momentum loss due to stellar winds is generally thought to be responsible for the Skumanich (\cite{Skumanich72}) law, but the exact dependence of rotational velocity on age is not entirely clear, and it relies on the assumed stellar magnetic field geometry and degree of core-envelope coupling (Kawaler \cite{Kawaler88}; Krishnamurthi et al. \cite{Krishnamurthi97}). Empirical age-rotation relations have been proposed to determine the age of stars, a method referred to as gyrochronology  (e.g., Barnes \cite{Barnes03}, \cite{Barnes07}; Mamajek \& Hillenbrand \cite{Mamajek08}; Collier Cameron et al. \cite{CollierCameron09}; Barnes \cite{Barnes10}). Empirical age-activity relations have also been proposed, though these do not always find activity decaying with time quite as simply as predicted by the Skumanich law (e.g., Feigelson et al. \cite{Feigelson04}; Pace \& Pasquini \cite{Pace04}; Giampapa et al. \cite{Giampapa06}).\\
\indent
The determination of stellar rotational periods in large samples of stars of different ages and mass is also crucial for understanding the relationships with stellar properties and their close surroundings. It is also expected that the dissipation timescale of circumstellar discs is related to the processes of planets formation (e.g. Bouvier \cite{Bouvier08}). On the other hand, the presence of planets may in turn influence the stellar rotational evolution. Recently, Pont (\cite{Pont09}) has  investigated the influence of tidal interaction with close-in giant planets (Hot Jupiters), and Lanza (\cite{Lanza10}) studied the role of Hot Jupiter interaction with the coronal magnetic field in the stellar angular momentum loss rate via a magnetised wind.\\
\indent
Open clusters also represent a unique testbed for studying the Li depletion with age and its relationships with stellar rotation (e.g., Sestito \& Randich \cite{Sestito05}, Jeffries \& Oliveira \cite{Jeffries05}, Randich et al. \cite{Randich05}). In fact, stellar rotation plays a role in shaping the internal circulation, which in turn affects the abundance of light elements that are easily destroyed in the stellar interior (Li, Be, B). A survey of Li abundances in young stellar associations has been carried out by da Silva et al. (\cite{daSilva09}), who also studied the relationships with age and $v\sin i$. \\
\indent
While significant progress has been attained in the last few years, thanks to several dedicated projects or as a by-product of high-precision photometric planet-transit surveys, our knowledge of the rotation evolution of late-type stars remains incomplete, so firm confirmation of the correlations proposed above is still needed. RACE-OC (Rotation and ACtivity Evolution in Open Clusters) is a long-term project designed to study the evolution of the rotational properties and the magnetic activity of late-type members of stellar open clusters with ages between 1 to about 600 Myr (Messina \cite{Messina07}; Messina et al.  \cite{Messina08},  \cite{Messina10}). In Messina et al. (\cite{paper1}, hereafter Paper I), we considered stellar associations at distances closer than 100 pc and ages younger than about 100 Myr from the list of Torres et al. (\cite{Torres08}). These are TW Hydrae, $\beta$ Pictoris, Tucana/Horologium, Columba, Carina, and AB Doradus associations. They span ages between 8 and ∼100 Myr and, therefore, allow us to study the angular momentum evolution close to the crucial phase of dissipation of circumstellar discs and planet formation. In Paper I, we determined rotational periods for 144 of 204 late-type members of these associations. A clear indication of evolution with time of rotational period and of its photometric signatures (e.g., photometric amplitude of light curve) has been observed. The goal of the present paper is to complete the study of the associations identified by Torres et al. (\cite{Torres08}), determining the rotational properties of members of $\epsilon$̨ Chamaeleontis, Octans, and Argus associations ($\sim$6, 20, and 40 Myr, respectively), with mean distances between 110 and 150 pc.\\
\indent
In Sect.2, we present the sample considered in the present study. In Sect. 3, we describe the photometric data that are used in our analysis. In Sect. 4, we describe our procedure for the rotation period search. In Sect. 5, we present our results. Conclusions are given in Sect. 6.   In the Appendix we present the results on periods for a few revisited targets in Paper I. 

\rm

\section{The sample}

The sample of our investigation is taken from the compilation of Torres et al. (\cite{Torres08}), which
includes an updated analysis of the membership of nearby associations younger than 100 Myr.
We selected the following associations that have  mean distances beyond
 100 pc: {\object{$\epsilon$ Chamaeleontis},  \object{Octans}, and \object{Argus}.
These associations are reported with ages in the range  $\sim$6 to $\sim$40 Myr (cfr. Table\,\ref{tab-list}).

 The \object{$\epsilon$ Chamaeleontis} association was discovered and characterised by  Frink et al. (\cite{Frink98}). \rm
We selected 15 late-type members:  14 are high-probability members from the compilation of Torres et al. (\cite{Torres08}) and one is a recently added
member by Kiss et al. (\cite{Kiss11}) as part of the RAVE (Radial Velocity Experiment) project.
Given the possible connection with $\epsilon$ Chamaeleontis, as first suggested by Mamajek et al. (\cite{Mamajek99}),
we also compiled a list of all (15) late-type members of the $\eta$ Chamaeleontis cluster  from Torres et al. (\cite{Torres08}),
whose age is estimated as around 4-9 Myr. Four cluster members with good kinematic data were found 
as high-probability members of the $\epsilon$ Cha association.
In the following analysis we  consider both $\eta$ and $\epsilon$ Cha members as coeval with an age of $\sim$ 6 Myr, 
but we continue to use different symbols to distinguish them.\\
\indent
The Octans association was discovered within the SACY project (Torres et al. \cite{Torres03}). 
Owing to its mean distance of about 140 pc  and the small number of members with trigonometric parallax, the accuracies on
distance of individual members and age ($\sim$ 20 Myr) are poorer than for other associations
found with the same approach and the discovered members mostly have G spectral type.
Torres et al. (\cite{Torres08}) give an updated list of members of this association from which we selected a sample
of 12 members with spectral type later than F.

The Argus association was also discovered by Torres et al. (\cite{Torres03})  in the SACY survey. Based on the convergence method and 
following the suggestion of Makarov \& Urban (\cite{Makarov00}), they found that the kinematic properties of the members of the  IC\,2391 open cluster
are in good agreement with their proposed Argus members. Torres et al. (\cite{Torres08}) give an updated list of members of
both Argus and IC\,2391 members. Finally, we also considered the new candidates members of young associations recently identified by
Kiss et al.~(\cite{Kiss11}) from RAVE data  and by Desidera et al. (\cite{Desidera11}). 
From these lists, we compiled a sample of
57 members (27 Argus stars and 30 stars in IC\,2391) with spectral type later than F. \rm

From the initial list of  121 \rm known stellar members, we selected
 99 \rm late-type stars (spectral types later than F or colours consistent with
a late spectral type) that are 
suitable for the photometric search of rotational modulations.
In Table\,\ref{tab-list} we list name, abbreviation, age, and mean distance of the associations under study (Torres et al. \cite{Torres08}), together
with the number of known members and late-type members selected for period search.
Most of the spectroscopic information (spectral types, projected
rotational velocity $v\sin i$) is from SACY database
(Torres et al.~\cite{Torres06}). In the case of IC\,2391 cluster, $v\sin i$ values come from da Silva et al. (\cite{daSilva09}) and Marsden et al. (\cite{Marsden09}).
Additional bibliography
for individual targets is given in Appendix A.

Furthermore, to complete the analysis of the associations studied in Paper I, we considered
the photometry becoming available in the Wide Angle Search for Planets (SuperWASP) archive for the stars of TW Hydrae, $\beta$ Pictoris, Tucana/Horologium, Columba, Carina 
and AB Doradus associations.  The new rotation periods of the associations studied in Paper I and retrieved from SuperWASP photometry are reported in Appendix B. \rm

\begin{table*}
\scriptsize
\caption{\label{t:eps Cha} {\bf  $\epsilon$ Cha association and $\eta$ Cha cluster}: summary of period searches.}
\centering
\begin{tabular}{llll@{\hspace{.1cm}}c@{\hspace{.1cm}}c@{\hspace{.1cm}}c@{\hspace{.1cm}}c@{\hspace{.1cm}}l@{\hspace{.1cm}}llll@{\hspace{.1cm}}}
\hline
 Name   &  ID$_{\rm ASAS}$   &P  &  $\Delta$P   & Timeseries & Qual. & $\Delta$V$_{\rm max}$ & $\sigma_{\rm acc}$ & V$_{min}$ & B$-$V & Sp.T. & Note & Variable's  \\  
           &                                 & (d)  &  (d)      &   Sections       & Flag   & (mag)     & (mag)              & (mag) & (mag) &  & & Name   \\  
\hline
\multicolumn{13}{c}{}\\
\multicolumn{13}{c}{\bf $\epsilon$ Chamaeleontis}\\
\multicolumn{13}{c}{}\\
      \object{CP-68\,1388}  &  105749-6914.0 &  3.560     &  0.030 &  12(13)/15 & C & 0.25 &   0.03 & 10.38 &  0.86$^{Av}$ & K1V(e)     &  new  &      \\
\object{GSC\,9419-01065}   &  114932-7851.0 &  8.000$^{c}$ &  0.100 &   4(10)/12 & U & 0.37 &   0.03 & 12.61 &  1.27$^{Av}$ & M0Ve     &  new  &  DZ\,Cha    \\
       \object{HIP\,58285}  &  115715-7921.5 &   ...      &   ...  &        ... &  ...  &   ... & ... & 12.0  &  0.93$^d$ & K0e     &  ...  &   T\,Cha   \\
        \object{HIP\,58490} &  115942-7601.4 &  8.000     &  0.100 &   7(11)/13 & C & 0.20 &   0.03 & 11.02 &  1.11$^{Av}$ & K4Ve     &  new  &      \\
       \object{HD\,104237E} &  120005-7811.6 &  2.45      &  ...  &         1/1 &  U &  ... &   ...  & 12.08 &  1.11$^d$ & K4Ve     &   B;   P$_{\rm lit}$ &   \\
        \object{HD\,104467} &  120139-7859.3 &  4.430     &  0.040 &    4(8)/13 & L & 0.05 &   0.03 &  8.45 &  0.63$^{Av}$ & G3V(e)     &  new  &      \\
    \object{GSC\,9420-0948} &  120204-7853.1 &  4.450     &  0.040 &  11(13)/13 & C & 0.48 &   0.03 & 12.30 &  1.00$^{Av}$ & M0e     &  new  &      \\
    \object{GSC\,9416-1029} &  120437-7731.6 &  5.350 &  ... &          1/1 &  U & ... &    ... & 13.81 &  1.30$^{Av}$ & M2e     &  SB2; P$_{\rm lit}$  &      \\
        \object{HD\,105923} &  121138-7110.6 &  5.050     &  0.050 &  11(13)/14 &  C & 0.07 &   0.03 &  9.07 &  0.73$^{Av}$ & G8V     &  new  &      \\
    \object{GSC\,9239-1495} &  121944-7403.9 &   ...      &   ...  &        ... &  ...  & ... &   ... & 12.54 &  1.28$^{Av}$ & M0e     &  ...  &      \\
    \object{GSC\,9239-1572} &  122023-7407.7 &  1.536      &  0.002 &      9(8)/9 & C & 0.30 &   0.03 & 12.85 &  0.97$^{Av}$ & M1V     & V; new  &      \\
       \object{CD-74\,712}  &  123921-7502.7 &  3.970     &  0.030 &  23(24)/25 & C & 0.13 &   0.03 & 10.22 &  0.97 & K3Ve     &  new  &      \\
 \object{GSC\,9235-1702}  &  122105-7116.9 &  6.85      & 0.05   &   9(14)/14 & C & 0.25 &   0.02 & 11.62 & 1.20$^d$  & K7V      & P=P$_{\rm lit}$ & \\ 
      \object{CD-69\,1055}  &  125826-7028.8 &  2.007     &  0.007 &   8(13)/15 & C & 0.10 &   0.03 &  9.91 &  0.83$^{Av}$ & K0Ve     &  new  &      \\
 \object{TYC\,9246-971-1}   &  132208-6938.2 &  3.750     &  0.030 &   9(14)/15 & C & 0.13 &   0.03 & 10.31 &  0.86$^{Av}$ & K1Ve     &  new  &  MP\,Mus     \\

\hline
\multicolumn{13}{c}{}\\
\multicolumn{13}{c}{\bf $\eta$ Chamaeleontis cluster}\\
\multicolumn{13}{c}{}\\
    \object{RECX\,1}     &  083656-7856.8    & 4.50 &  0.040 &  12(12)/15 & C & 0.13  & 0.03 & 10.36 &  1.19 & K4V      & V; P=2P$_{\rm lit}$    &  EG\,Cha  \\
    \object{RECX\,3}$^a$ &  084136-7903.7    &  3.57  & ...    & 2/2        & U & 0.08  & ...  & 14.30 & ...   & M3       & P$_{\rm lit}$     &  EH\,Cha  \\
    \object{RECX\,4}     &  084229-7903.9    &  7.1   &   0.4  &  6(10)/12  & C & 0.25  & 0.03 & 12.73 & ...   & K7       & P=P$_{\rm lit}$   &  EI\,Cha  \\
    \object{RECX\,5}     &  084235-7858.1    &  3.11  & ...    & 1/2        & U & 0.03  & ...  & 15.17 & ...   & M5       & P$_{\rm lit}$     &  EK\,Cha  \\
    \object{RECX\,6}$^a$ &  084239-7854.7    &  1.84  & ...    & 2/2        & U & 0.16  & ...  & 13.98 & ...   & M7       & P$_{\rm lit}$     &  EL\,Cha  \\
    \object{RECX\,7}$^a$ &  084302-7904.7    &  2.62  & ...    & 2/2        & U & 0.09  & ...  & 10.76 & ...   & M3       & SB2; P$_{\rm lit}$     &  EM\,Cha  \\
    \object{RECX\,9}$^a$ &  084419-7859.1    &  1.71  & ...    & 2/2        & U & 0.38  & ...  & 14.75 & ...   & M4       & V; P$_{\rm lit}$     &  EN\,Cha  \\
    \object{RECX\,10}    &  084432-7846.6    &  20.0  & 3.0    &  5(5)/12   & C & 0.26  & 0.03  & 12.41 & ...   & K7       & P=P$_{\rm lit}$     &  EO\,Cha  \\
    \object{RECX\,11}    &  084708-7859.6    &  4.84  & 0.04   &  7(7)/12   & C & 0.28  & 0.03  & 11.00 & ...   & K4       & P$\ne$P$_{\rm lit}$     &  EP\,Cha  \\
    \object{RECX\,12}    &  084802-7855.0    &  1.26$^b$  &  0.05  &    2(2)/12 & C & 0.18  & 0.03 & 13.92 &  0.73 & M3e   & V; P= P$_{\rm lit}$  &  EQ\,Cha  \\
    \object{RECX\,14}    &  ...              &  1.73  & ...    &  2/2      & U & 0.04 &  ...  & 17.07 & ...   & M5e      & P$_{\rm lit}$     &  ES\,Cha  \\ 
    \object{RECX\,15}$^a$ &  084319-7905.4    & 12.8   & ...    &  2/2       & U & 0.44  & ...  & 13.97 & ...   & M3e      & P$_{\rm lit}$     &  ET\,Cha  \\ 
    \object{RECX\,16}    &.....              & ...    & ...    &  ...       & ....  & ... & ... & 17.07  & ...   & M5e      & ...               &  \\
    \object{RECX\,17}    &......             & ...    & ...    &  ...       & ....  & ...  & ...& 16.82 & ...   & M5e      & ...               &  \\
    \object{RECX\,18}    &......             & ...    & ...    &  ...       & ....  & ...  & ...& 17.66 & ...   & M5e      & ...               &  \\
\hline
\multicolumn{13}{l}{}\\
\multicolumn{13}{l}{V=visual companion; B=binary system; P$_{\rm lit}$: period as given in the literature; $^a$: ASAS very sparse observations;}\\
\multicolumn{13}{l}{ $^b$: period undetected in the periodogram of the complete time series; $^c$: $v\sin i$ inconsistent with v$_{\rm eq}$=2$\pi$R/P};\\
\multicolumn{13}{l}{$^d$: from Sp.\,type; $^{Av}$: V mag and B$-$V corrected for reddening.}
\end{tabular}
\end{table*}

\begin{table*}
\scriptsize
\caption{\label{t:octans} As in Table\,\ref{t:eps Cha}  for {\bf Octans Association}.}
\centering
\begin{tabular}{llll@{\hspace{.1cm}}c@{\hspace{.1cm}}c@{\hspace{.1cm}}c@{\hspace{.1cm}}c@{\hspace{.1cm}}l@{\hspace{.1cm}}llll@{\hspace{.1cm}}}
\hline
  Name   &  ID$_{\rm ASAS}$   &P  &  $\Delta$P   & Timeseries & Qual. & $\Delta$V$_{\rm max}$ & $\sigma_{\rm acc}$ & V$_{min}$ & B$-$V & Sp.T. & Note & Variable's  \\   
            &                                 & (d)  &  (d)      &   Sections       & Flag  & (mag)     & (mag)              & (mag) & (mag) &  & & Name   \\   
\hline
       \object{CD-58\,860} &  041156-5801.8 &  1.612 &  0.005 & 9(10)/12 & C &  0.10 &   0.03 &  9.92 &  0.68 & G6V     &  new  &      \\
       \object{CD-43\,1451} &  043027-4248.8 & ...   &   ...   &  ...   &  ...   &   ... & ...  & 10.75 &  0.79 & G9V(e)     &  ...  &      \\
       \object{CD-72\,248} &  050651-7221.2 &  0.236 &  0.001 & 6(6)/12 & C & 0.09 &   0.03 & 10.70 &  0.82 & K0I     &  new  &      \\
       \object{HD\,274576} &  052851-4628.3 &  2.220 &  0.020 & 9(10)/12 & C &  0.10 &   0.03 & 10.48 &  0.66 & G6V     &  new  &      \\
       \object{CD-47\,1999} &  054332-4741.2 &  1.265 &  0.004 & 3(6)/12 & C & 0.08 &   0.03 & 10.05 &  0.56 & G9V(e)     &  P=P$_{\rm ACVS}$  &      \\
       \object{TYC\,7066-1037-1} &  055812-3500.8 &  2.470$^b$ &  0.040 & 6(9)/14 & C & 0.08 &   0.03 & 11.14 &  0.80$^{Av}$ & G9V     &  new  &      \\
       \object{CD-66\,395} &  062515-6629.5 &  0.371 &  0.001 & 6(10)/12 & C & 0.16 &   0.03 & 10.75 &  0.71 & K0I     &  new  &      \\
       \object{TYC\,9300-0529-1} &  184949-7157.2 &  2.550$^b$ &  0.030 & 4(14)/14 & L & 0.06 &   0.03 & 11.59 &  0.80 & K0V     &  V; new  &      \\
       \object{TYC\,9300-0891-1} &  184949-7157.2 &   ...  &   ...   &   ...    &  ...   &   ... & ... & 11.43 &  0.77 & K0V(e)     &  V  &      \\
       \object{CP-79\,1037} &  194704-7857.7 &  1.950$^b$ &  0.020 & 5(6)/13 &  C & 0.05 &   0.03 & 11.18 &  0.75 & G8V     &  new  &      \\
       \object{CP-82\,784} &  195357-8240.7 &  1.373 &  0.003 & 9(9)/13 & C & 0.08 &   0.03 & 10.81 &  0.85 & K1V     &  new  &      \\
       \object{CD-87\,121} &  235818-8626.4 &  0.762 &  0.002 & 4(6)/12 & L & 0.03 &   0.03 &  9.94 &  0.74 & G8V     &  V?; new  &      \\
\hline
\multicolumn{13}{l}{}\\
\multicolumn{13}{l}{V=visual companion; P$_{\rm ACVS}$: period in ACVS; $^b$: period undetected in the periodogram of the complete time series;}\\
\multicolumn{13}{l}{$^{Av}$: V mag and B$-$V corrected for reddening.}
\end{tabular}

\end{table*}

\begin{table*}
\scriptsize
\caption{\label{t:argus}  As in Table\,\ref{t:eps Cha}  for  {\bf Argus Association and IC\,2391 cluster}.}
\centering
\begin{tabular}{llll@{\hspace{.1cm}}c@{\hspace{.1cm}}c@{\hspace{.1cm}}c@{\hspace{.1cm}}c@{\hspace{.1cm}}l@{\hspace{.1cm}}l@{\hspace{.1cm}}l@{\hspace{.1cm}}l@{\hspace{.1cm}}l@{\hspace{.1cm}}}
\hline
  Name   &  ID$_{\rm ASAS}$   &P  &  $\Delta$P   & Timeseries & Qual. &$\Delta$V$_{\rm max}$ & $\sigma_{\rm acc}$ & V$_{min}$ & B$-$V & Sp.T. & Note & Variable's  \\   
            &                                 & (d)  &  (d)      &   Sections       & Flag & (mag)     & (mag)              & (mag) & (mag) &  & & Name   \\   
\hline
\multicolumn{13}{c}{}\\
\multicolumn{13}{c}{\bf Argus}\\ 
\multicolumn{13}{c}{}\\
          \object{HD\,5578}    &  005655-5152.5 &  1.461 &  0.005 &  4(5)/6 & L & 0.05 &   0.03 &  9.62 &  0.99 & K3V & V; new & BW\,Phe \\
          \object{CD-49\,1902} &  054945-4918.4 &  1.100 &  0.005 &  3(6)/8 & L & 0.09 &   0.03 & 11.30 &  0.69 & G7V & new &  \\
          \object{CD-56\,1438} &  054945-4918.4 &  0.240 &  0.002 &  1(4)/8 & U & 0.08 &   0.03 & 11.10 &  0.66 & K0V & new &    \\
          \object{CD-28\,3434} &  064945-2859.3 &  3.820 &  0.030 & 7(7)/10 & C & 0.06 &   0.03 & 10.38 &  0.75$^a$ & G7V & new &    \\
          \object{CD-42\,2906} &  070153-4227.9 &  3.950 &  0.060 &  4(6)/7 & L & 0.09 &   0.03 & 10.50 &  0.84 & K1V & new &    \\
          \object{CD-48\,2972} &  072822-4908.6 &  1.038 &  0.004 &  6(3)/8 & C & 0.11 &   0.03 &  9.78 &  0.80 & G8V & P$\ne$P$_{ACVS}$   &  \\
\object{HD\,61005}              &  073547-3212.2 &  5.04   &  0.03   &  3(10)/14 & C & 0.05 & 0.03 & 8.19 & 0.75 & G8V & new  & \\
          \object{CD-48\,3199} &  074726-4902.9 &  2.190 &  0.010 &  6(5)/7 & C & 0.06 &   0.03 & 10.40 &  0.65 & G7V & new  &   \\
          \object{CD-43\,3604} &  074851-4327.3 &  0.890 &  0.001 &  4(6)/9 & L & 0.10 &   0.03 & 10.88 &  0.92 & K4V & new  &   \\
     \object{TYC\,8561-0970-1} &  075355-5710.1 & ...    & ...    &    ...  &  ... & 0.00  &   0.00 & 11.50 &  0.83$^{Av}$ & K0V & ...  &   \\
          \object{CD-58\,2194} &  083912-5834.5 &  5.160$^c$ &  0.040 &  3(4)/7 & U & 0.04 &   0.03 & 10.08 &  0.62 & G5V & new  &   \\
          \object{CD-57\,2315} &  085008-5746.0 & ...    & ...    &     ... &  ...  & ... &   ...   & 10.21 &  0.83 & K3V & ...  &   \\
     \object{TYC\,8594-0058-1} &  090204-5808.8 &  0.982 &  0.002 &  4(4)/7 & U & 0.05 &   0.03 & 11.08 &  0.71 & G8V & new  &   \\
          \object{CPD-62\,1197} &  091330-6259.2 &  1.260$^c$ &  0.001 &  7(5)/9 & U & 0.05 &   0.03 & 10.46 &  0.81 & K0V & new  &   \\
     \object{TYC\,7695-0335-1} &  092854-4101.3 &  0.391 &  0.001 &  7(7)/9 & C & 0.10 &   0.03 & 11.62 &  0.67 & K3V & new  &   \\
            \object{HD\,84075} &  093618-7820.7 & ...    & ...    &     ... &  ...  &   ...  & ... &  8.59 &  0.59 &  G1 & ...  &   \\ 
     \object{TYC\,9217-0641-1} &  094247-7239.8 &  2.300 &  0.010 &  6(6)/8 & C & 0.16 &   0.03 & 12.18 &  0.65 & K1V & new  &   \\
          \object{CD-39\,5883} &  094718-4002.9 &  4.060$^b$ &  0.030 &  5(5)/9 & C & 0.04 &   0.03 & 10.74 &  0.81$^a$ & K0V & new  &   \\
           \object{HD\,85151A} &  094843-4454.1 &  0.970 &  0.005 &  5(5)/5 & C & 0.02 &   0.01 &  9.10 &  0.50 & G7V & V; new  &   \\
           \object{CD-65\,817} &  094909-6540.3 &  2.740 &  0.020 &  4(4)/7 & U & 0.04 &   0.03 & 10.33 &  0.63 & G5V & V; new  &   \\
           \object{HD\,309851} &  095558-6721.4 &  1.816 &  0.007 &  5(5)/7 & C & 0.06 &   0.03 &  9.78 &  0.60 & G1V & new  &   \\
           \object{HD\,310316} &  104956-6951.4 &  3.610 &  0.030 &  4(6)/9 & L & 0.06 &   0.03 & 10.82 &  0.73 & G8V & V; new  &   \\
          \object{CP-69\,1432} &  105351-7002.3 &  1.030 &  0.002 &  3(5)/8 & L & 0.06 &   0.03 & 10.66 &  0.62 & G2V & new  &   \\
           \object{CD-74\,673} &  122034-7539.5 &  3.480 &  0.020 &  6(6)/8 & C & 0.07 &   0.03 & 10.49 &  1.02 & K3V & SB; new  &   \\
           \object{CD-75\,652} &  134913-7549.8 &  2.270 &  0.010 &  9(8)/9 & C & 0.08 &   0.03 &  9.61 &  0.68 & G1V & new   &  \\
          \object{HD\,133813}  &  151223-7515.3 &  4.090 &  0.030 & 8(7)/10 & C & 0.07 &   0.03 &  9.27 &  0.84 & G9V & new  & NY\,Aps \\
          \object{CD-52\,9381} &  200724-5147.5 &  0.830 &  0.001 &  5(6)/9 & C & 0.12 &   0.03 & 10.53 &  1.24 & K6V & P=P$_{ACVS}$  &   \\
\hline
\multicolumn{13}{c}{}\\
\multicolumn{13}{c}{\bf IC\,2391}\\
\multicolumn{13}{c}{}\\
\hline
            PMM\,7422\,\,\, \object{TYC 8162-1020-1}  &  082844-5205.7 & 1.512  &  0.005 &  7(6)/9 &  C & 0.11 &   0.03 & 10.40 &  0.69 &  G6     & new   &  \\
            PMM\,7956\,\,\, \object{1RXS J082952.7-514030}&  082952-5140.7 & ...    & ...    &     ... &  ... & 0.00 &    0.00 & 11.50 &  0.97 &  K2e     &  ...  &      \\
            PMM\,1560\,\,\, \object{TYC 8568-1413-1}&  082952-5322.0 & ...    & ...    &     ... &  ... &0.00 &    0.00 & 10.64 &  0.61 &  G3     &  ...  &      \\
            PMM\,6974\,\,\, \object{TYC 8162-754-1} &  083418-5216.0 &  7.800 &  0.050 &  4(3)/8 &  U & 0.10 &   0.03 & 12.04 &  1.04 &  ...    &  new  &      \\
            PMM\,4280\,\,\, \object{TYC 8568-407-1}  &  083421-5250.1 & ...    & ...    &     ... &  ... &0.00 &    0.00 & 10.04 &  0.67 &  G5     &  ...  &      \\
            PMM\,6978\,\,\, \object{1RXS J083502.7-521339}  &  083501-5214.0 &  5.100 &  0.200 &  4(5)/8 &  L & 0.11 &   0.03 & 12.01 &  1.02 &  ...    &  new  &      \\
            PMM\,2456\,\,\, \object{1RXS J083543.4-532123}  &  083544-5321.3 & ...    & ...    &     ... & ... & 0.00 &    0.00 & 12.16 &  0.92 & K3e     &  ...  &      \\
            PMM\,351 \,\,\,\,\,\,\object{1RXS J083624.5-540101}  &  083624-5401.1 &  1.923$^c$ &  0.007 &  6(4)/7 & U & 0.06 &   0.03 & 10.17 &  0.57 &  G0     &  new  &      \\
            PMM\,3359  &  083655-5308.5 &  3.840 &  0.030 &  6(3)/8 & C & 0.05 &   0.03 & 11.38 &  0.76 &  ...    &  new  &      \\
            PMM\,5376 &  083703-5247.0 & ...    & ...    &     ... &  ... &0.00 &    0.00 & 13.94 &  1.37 &  Me     &  ...  &      \\
            PMM665 \,\,\,\,\,\, \object{2MASS J08375156-5345458}  &  083752-5345.8 &  4.510 &  0.040 &  3(4)/8 & U & 0.07 &   0.03 & 11.33 &  0.75 &  G8     &  new  &      \\
            PMM\,4336\,\,\, \object{VXR PSPC 2a}  &  083756-5257.1 &  3.71  &  0.03  &  4(5)/9 & L & 0.06 &   0.03 & 11.25 &  0.86 &  G9     &  new  &      \\
            PMM\,4362\,\,\, \object{CD-52 2467}  &  083823-5256.8 &  3.930 &  0.030 &  4(4)/9 & U & 0.05 &   0.03 & 10.91 &  0.67 &  ...     &  new  & VXR\,3a     \\
            PMM\,4413\,\,\, \object{CD-52 2472}  &  083856-5257.9 &  5.05 &  0.05 &  3(2)/9 & U & 0.03 &   0.03 & 10.20 &  0.67 &  G2     &  SB2, new  &  VXR\,5    \\
            PMM\,686\,\,\,\,\,\, \object{1RXS J083921.9-535420}  &  083923-5355.1 &  4.410 &  0.050 &  2(0)/7 & U & 0.05 &   0.03 & 12.55 &  1.04 &  K4e     &  new  &      \\
	    PMM\,4467\,\,\, \object{VXR PSPC 12}  &  083953-5257.9 &  3.850 &  0.030 &  7(7)/9 & C & 0.17 &   0.03 & 11.84 &  0.84 & K0(e    &  P=P$_{lit}$  & V364 Vel   \\
            PMM\,1083\,\,\, \object{VXR PSPC 14}  &  084006-5338.1 &  1.333 &  0.004 &  8(8)/8 & C & 0.07 &   0.03 & 10.38 &  0.57 &  G0     &  P=P$_{lit}$  &  V365 Vel    \\
            PMM\,8415\,\,\, \object{VXR PSPC 16a}  &  084016-5256.5 & ...    & ...    &     ... & ... & 0.00 &    0.00 & 11.63 &  0.87 & G9(e)   &  ...  &     \\
            PMM\,1759\,\,\, \object{VXR PSPC 18}  &  084018-5330.4 & ...    & ...    &     ... & ... & 0.00 &    0.00 & 12.54 &  1.01$^{Av}$ & K3e     &  ...  &      \\
            PMM\,1142\,\,\, \object{VXR PSPC 22a}  &  084049-5337.8 & ...    & ...    &     ... & ... & 0.00 &    0.00 & 11.04 &  0.68 &  G6     &  ...  &       \\
            PMM\,1820\,\,\, \object{VXR PSPC 35a}  &  084126-5322.7 &  0.527$^b$ &  0.001 &  4(5)/8 & C & 0.12 &   0.03 & 12.41 &  1.00 & K3e     &  P=P$_{lit}$  &  V366 Vel   \\
            PMM\,4636\,\,\, \object{VXR PSPC 41} &  084158-5252.2 & ...    & ...    &     ... &  ... &0.00 &    0.00 & 13.20 &  1.26 & K7e     &  ...  &   V368 Vel    \\
            PMM\,3695\,\,\, \object{VXR PSPC 47} &  084219-5301.9 &  0.223 & ...    &     ... & U & 0.00   &  0.00 & 13.60 &  1.44 & M2e     &  P$_{lit}$  &   V371 Vel   \\
            PMM\,756 &  084300-5354.1 &  3.140 &  0.020 &  7(7)/7 & C & 0.13 &   0.03 & 11.06 &  0.68 &  G9     &  new  &      \\
            PMM\,2012\,\,\, \object{VXR PSPC 69}  &  084359-5333.7 &  2.210 &  0.010 &  4(4)/8 & U & 0.10 &   0.03 & 11.41 &  0.83 & K0(e)   &  new  &     \\
            PMM\,4809\,\,\, \object{VXR PSPC 70}  &  084405-5253.3 &  2.600 &  0.010 &  6(7)/9 &  C & 0.08 &   0.03 & 10.73 &  0.64 & G3(e)   &  new  &    V376 Vel	  \\
            PMM\,1373 &  084410-5343.6 &  5.380 &  0.060 &  1(2)/8 & U & 0.06 &   0.03 & 12.06 &  0.96 & ...     &  new  &      \\
            PMM\,5884\,\,\, \object{VXR PSPC 72}&  084426-5242.5 &  3.030 &  0.020 &  5(6)/9 & C & 0.07 &   0.03 & 11.35 &  0.73 & G9(e)   &  P=P$_{lit}$  &  V377 Vel  \\
            PMM\,4902\,\,\, \object{ VXR PSPC 76a} &  084526-5252.0 &  4.820$^b$ &  0.050 &  4(2)/9 & C & 0.10 &   0.03 & 12.45 &  1.05 & K3e     &  P=P$_{lit}$  &   V379 Vel    \\
            PMM\,2182\,\,\, \object{TYC 8569-2769-1} &  084548-5325.8 &  1.437$^b$ &  0.002 &  8(4)/8 & C & 0.06 &   0.03 & 10.18 &  0.63 & G2(e)   &  new  &      \\
\hline
\multicolumn{13}{l}{}\\
\multicolumn{13}{l}{V=visual companion; SB=binary system; P$_{\rm lit}$: period as given in the literature; P$_{\rm ACVS}$: period in ACVS; }\\
\multicolumn{13}{l}{$^b$: period undetected in the periodogram of the complete time series; $^c$: $v\sin i$ inconsistent with v$_{\rm eq}$=2$\pi$R/P;  }\\
 \multicolumn{13}{l}{$^{Av}$: V mag and B$-$V corrected for reddening.}\\
\end{tabular}

\end{table*}

\begin{figure*}[t,*h]
\begin{minipage}{18cm}
\centerline{
\includegraphics[trim=0 0 0 0,width=8.5cm,height=9.5cm,angle=90]{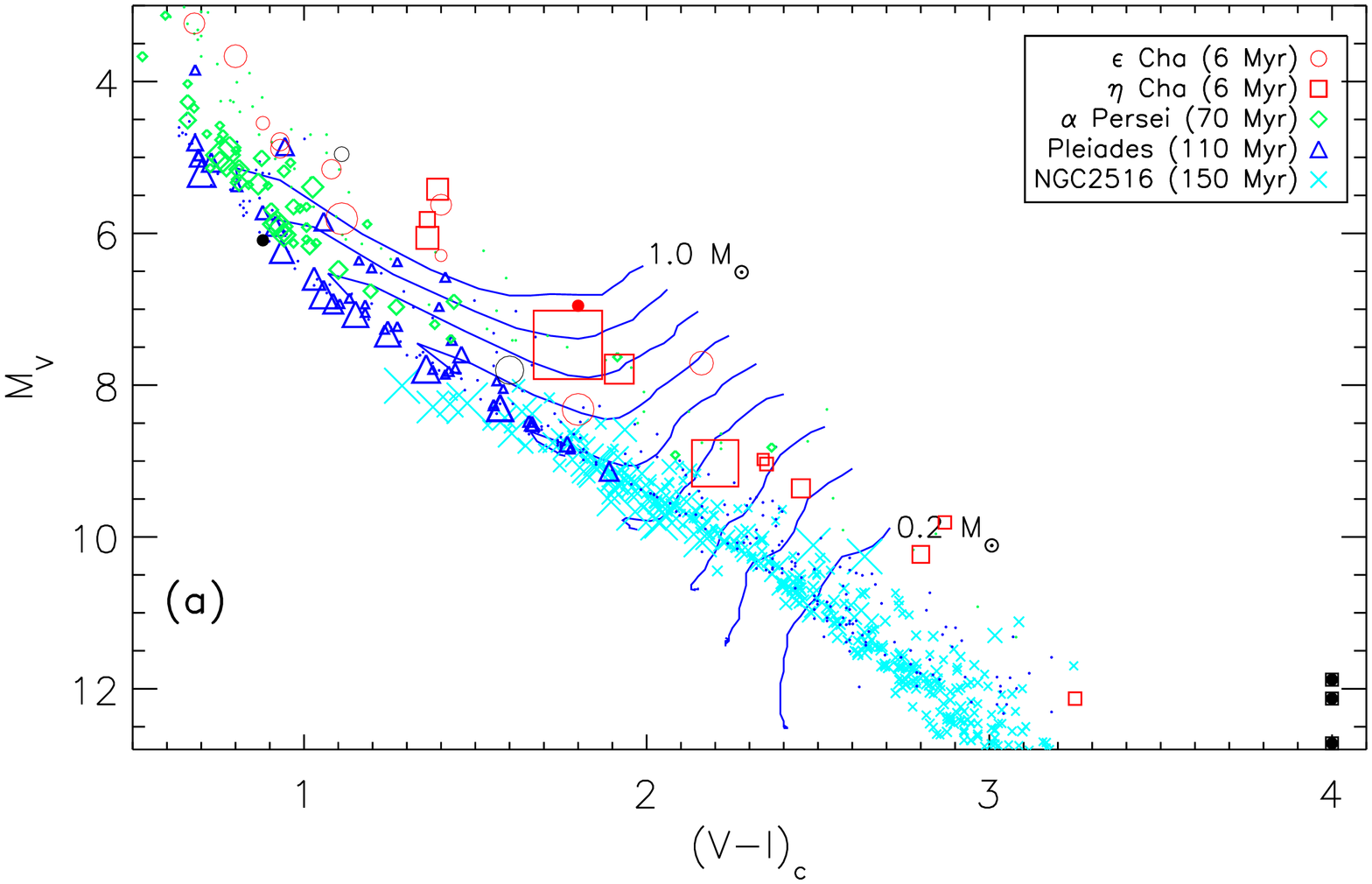}
\includegraphics[trim=0 0 0 0,width=8.5cm,height=9.5cm,angle=90]{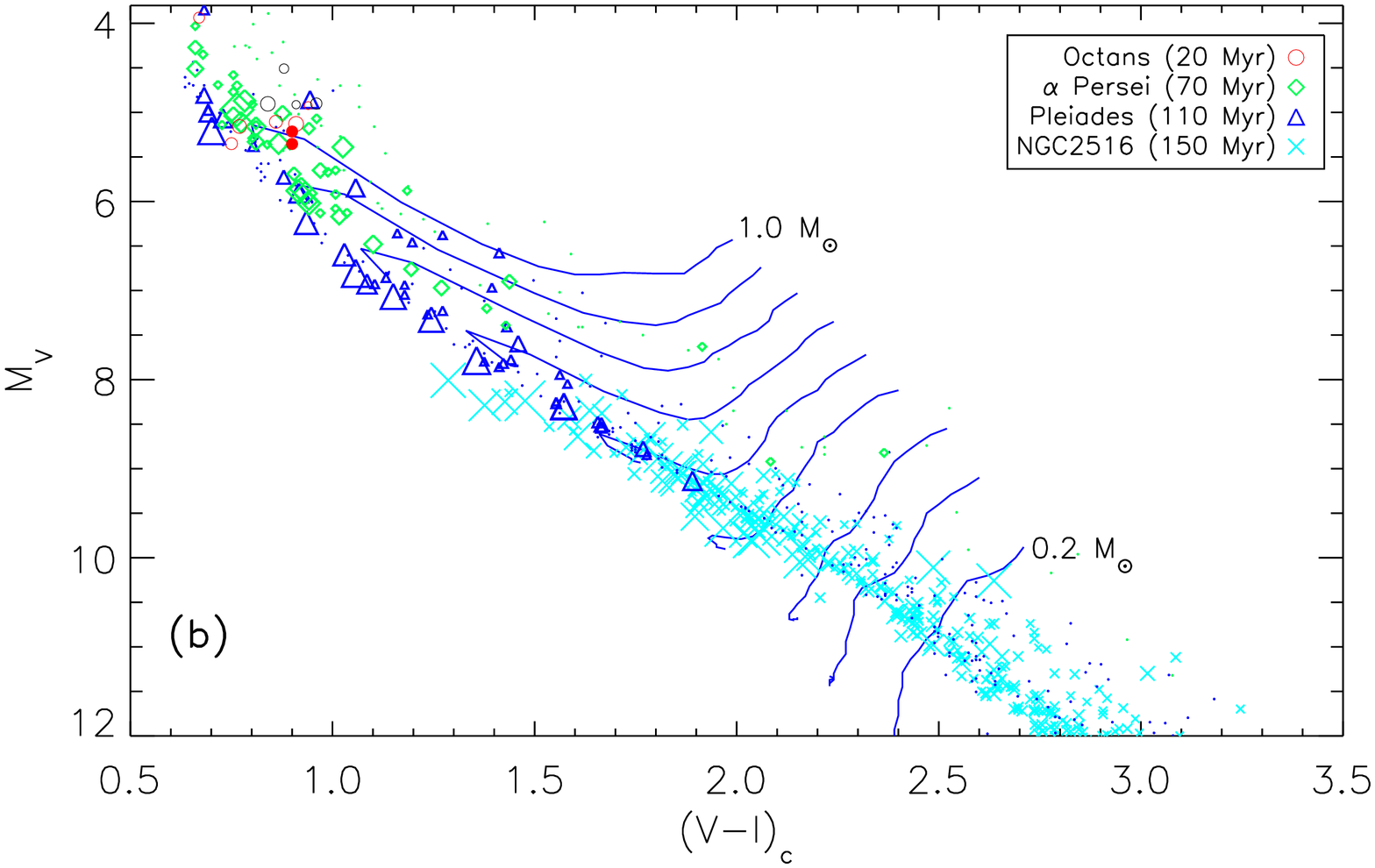}
}
\end{minipage}
\caption{ Colour-corrected \rm colour-magnitude diagrams with overplotted PMS tracks (solid lines) from Baraffe et al.\,(\cite{Baraffe98}) for $\epsilon$ and $\eta$ Cha (left) and Octans (right).
Different symbols indicate stars belonging to different association/clusters;  bigger symbols indicate longer rotation periods. 
Filled symbols represent members with unknown rotation period. Black symbols represent stars whose V$-$I colour is
derived from spectral type.\label{cmd}}
\end{figure*}

\section{Data}

\subsection{The ASAS photometry}

As in Paper I, most of the present analysis is based on data from the All Sky Automated Survey (ASAS) 
(Pojmanski \cite{Pojmanski97}; \cite{Pojmanski02}).
The ASAS project is monitoring all stars brighter than V=14 at declinations $\delta < +28^{\circ}$, 
with typical sampling of 2 days.
The ASAS archive\footnote{www.astrouw.edu.pl/asas/} ensures long-term monitoring since it contains data from 1997 to date. 
A short-term monitoring might hamper the identification of the correct rotational period. Since the configuration of active regions, which is
responsible for the light rotational modulation,  evolves with time, such a long time-span is particularly suitable for our purposes. \rm

The linear scale at focal plane is 16 arcsec/pixel. 
The FWHM of stellar images is 1.3-1.8 pixels. 
Aperture photometry through five apertures is available in the ASAS catalogue. The choice of the aperture
 adopted in our analysis was made star by star by selecting the aperture 
giving the highest photometric precision (i.e., the minimum average magnitude uncertainty).

\subsection{The SuperWASP photometry}

The SuperWASP project (Pollacco et al.~\cite{Pollacco06}) recently released the first public data archive 
(Butters et al. \cite{Butters10}) \footnote{www.wasp.le.ac.uk/public/}. Light curves from 2004 to  
2008 for both the northern and southern observatories are included.
Temporal and spatial coverage is irregular, but extended enough to make  a systematic search 
for photometric timeseries of our targets meaningful. 
The observing procedure includes sequences up to nine hours long for stars at the most favourable sky declinations with 
a sampling of about ten minutes. The SuperWASP data have therefore a better sensitivity to periods $\la$ 1 d than the ASAS data.
We then checked the availability in SuperWASP archive of 
data for both the targets studied in this paper as well those we studied in Paper I.
To further complete Paper I analysis, we also checked the availability of ASAS and SuperWASP photometry 
for the new candidates members of young MG identified by Kiss et al.~(\cite{Kiss11}).
Our analysis is based on the processed flux measurements obtained through application 
of the SYSREM algorithm (Tamuz et al. ~\cite{Tamuz05}). SuperWASP observations  were collected in 2004 
without any light filter,  the spectral transmission defined by the optics, detector, and atmosphere. Starting from 2006
they were collected through a wide band filter in the 400-700 nm. \rm        Owing to differences in spectral bands with respect to
the  standard Johnson V ASAS data, \rm we decided to analyse the SuperWASP data independently without merging these with ASAS data.

\subsection{Data from the literature}

Literature period determinations, including those listed in ASAS Catalogue of Variable Stars (ACVS) and the search
for variable stars from SuperWASP data (Norton et al.~\cite{Norton07}), were considered and
 compared to our measurements when available.
These are listed in Tables \ref{t:eps Cha}-\ref{t:argus}. Three cases of discrepant results
are discussed individually in Appendix A.

\section{Photometry rotation period search}

The rotational period search procedure is described in Paper I. Here we briefly summarise it.
The analysis makes use of the Lomb-Scargle periodogram (Scargle \cite{Scargle82}).
The light curves of late-type active stars are typically characterised by changes in amplitude and
shape with timescales of a few months or even shorter  (see, e.g., Messina et al. \cite{Messina04}). These
are due to the finite lifetime of active regions, consisting of either dark spots or bright faculae, and to differential rotation. Additional photometric
variability is observed on timescales shorter and longer than the rotation period because of
flares and magnetic cycles.
This variability pattern causes the application of the Lomb-Scargle periodogram
to the whole light curve to fail the detection of the rotational period in some cases. 
Therefore, in addition to analysing the whole timeseries (which is typically 8-yr long for ASAS data and 4-yr long for SuperWASP data), 
we performed the period search on light-curve segments not exceeding two months.

The determination of the false alarm probability (FAP) is a crucial issue for  evaluating  the
significance of detected periodicities. The existence of significant correlations between
data that are collected on shorter timescales  than the stellar variability does not allow
safely  making the assumption that each observed data point is
independent of the  others (Herbst  \& Wittenmyer \cite{Herbst96};  Stassun et  al. \cite{Stassun99}; 
Rebull \cite{Rebull01};  Lamm  et al. \cite{Lamm04}).
To overcome this difficulty, we follow the bootstrap approach proposed by Herbst et al.~(\cite{Herbst02}).
For each light curve segment, we scramble 1000 times the day numbers of  the 
Julian Day (JD) while keeping photometric
magnitudes  and the  decimal part  of  the JD  unchanged.
Then we perform the period search on each fake randomised dataset, comparing the
power of the highest peak to that observed on the real dataset to obtain the FAP.
We adopt a confidence level larger than  99\% (FAP$<0.01$) as detection threshold
for a significant detection.

Finally, to successfully identify the true rotational periodicity, it is necessary to consider 
aliases, which might have even higher power than the true period.
The spectral window function is inspected to disentangle aliases
(1-d peak, beat periods  between the star's rotation period (P) and the data 
sampling).
The uncertainties in the period determination were derived following Lamm et al.  (\cite{Lamm04}).

For each target we report (Tables \ref{t:eps Cha}-\ref{t:argus}) the detected rotation period, together with the number of segments 
in which such a period was derived, with a confidence level higher than 99\%. In some cases, however, the same period was found in 
other segments with a confidence level below 99\%, but the lightcurve still showed a clear rotational modulation with this very period. 
We therefore also report the number of segments in which the lightcurve, when phased with the detected rotation period, shows a 
smooth sinusoidal variation; in this case, the average residual in the sinusoidal fit of the light curve is smaller than the light-curve amplitude. 
The number of segments in which the period was found with a confidence level higher than 99\% and/or the number of segments in 
which the phased lightcurve show average residuals lower than its amplitude can be used to estimate the robustness of the 
period determination. When $v\sin i$ measurements are also available, a consistency check can be performed with the equatorial 
velocity, which can be derived from the rotational period and an estimate of the stellar radius. We use this information to 
classify the detected periods according to their reliability.

Periods derived in at least five segments, or in less than five but with an independent confirmation from the literature, and that  are consistent with $v\sin i$ will be referred to as \it confirmed \rm periods  (C). \rm Periods derived in less than five segments that are consistent with $v\sin i$ and that produce mean residuals in the sinusoidal fit less than the light-curve amplitude in at least five segments will be referred to as \it likely \rm  periods   (L). \rm Periods derived in less than five segments that produce mean residuals in the sinusoidal fit less than the light-curve amplitude in less than five segments, or with only 1-2 determinations in the literatures, or that are inconsistent with $v\sin i$ will be referred to as \it uncertain \rm periods  (U). \rm

\begin{figure}[]
\begin{minipage}{10cm}
\centerline{
\includegraphics[trim=0 0 0 0,width=8.5cm,height=9.5cm,angle=90]{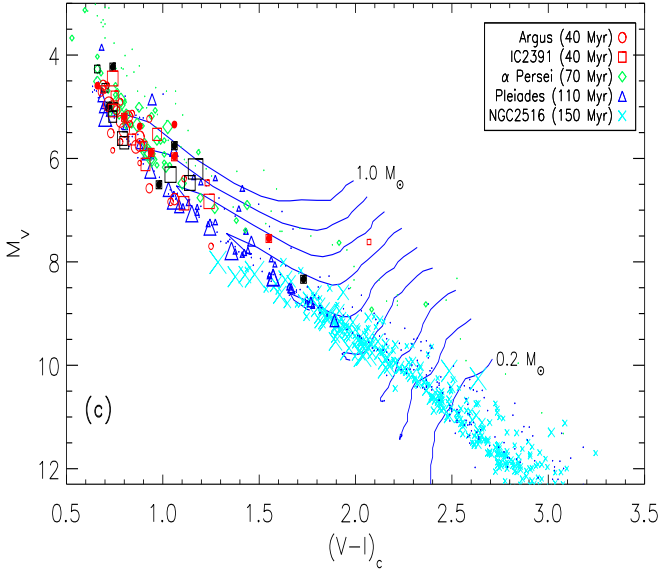}
}
\end{minipage}
\caption{As in Fig.\,\ref{cmd} for Argus and IC\,2391.\label{cmd1}}
\end{figure}

\begin{table}
\caption{Summary of results of the comparison between $v\sin i$ and equatorial velocity.\label{tab-vsini}}
\begin{tabular}{ccccc}
\hline
Association     &  \#		&	$<$sin$i$$>$$\pm$$\sigma$	&	r	&	Significance	\\
			& Stars	&                                             &              &   Level                  \\
\hline
$\epsilon$ Cha		&	10	&	0.85$\pm$0.22		&	0.89		& 	a	\\
$\eta$ Cha               &      8     &     0.80$\pm$0.25		&	0.61		&	b	\\
$\epsilon$/$\eta$ Cha	&	18	&	0.83$\pm$0.22		&	0.76		&	a	\\
Octans			&	10	&	0.79$\pm$0.22		&	0.94		&	a	\\
Argus			&	17	&	0.86$\pm$0.23		&	0.96		&	a	\\
IC2391			&	19	&	0.76$\pm$0.22		&	0.82		&	a	\\
Arg / IC2391		&	36	&	0.81$\pm$0.16		&	0.84		&	a	\\
\hline
\multicolumn{5}{l}{a: confidence level $>$ 99.999\%; b: confidence level = 99.995\%.}\\
\end{tabular}
\end{table}

\section{Results}

The results of our investigation are summarised in Table \ref{tab-list}, where we report the total number  of \it confirmed  \rm periodic members (and the total number of  \it confirmed \rm + \it likely \rm + \it uncertain \rm in paretheses), the number of periodic 
members with period adopted from the literature, new periods determined from this study (and periods revised by us  with respect to earlier literature values).

 We determined 45 \it confirmed  \rm rotation periods (31 of which are new determinations), 11 \it likely\rm, 24 \it uncertain \rm (10 of which are taken from the literature).  
Eleven of 80 periodic stars were already known from the literature and we could confirm their rotation period, whereas  we revised it for 3 stars (RECX\,1, RECX\,11, and CD-48\,2972,
 see Appendix A). The rotation periods of 19 stars remain unknown, for three of which we have data neither in ASAS,  in SuperWasp, nor in the literature, whereas for 16  we find non-periodic variability. \rm
In Tables \ref{t:eps Cha}-\ref{t:argus} we report the results of a period search in some detail  for the members of individual associations.
 As  in Paper I, the results of the period search for close spectroscopic binaries whose rotation might be altered
by the tides of their companions is here reported. However, they are excluded from the rotational
distributions and also when computing the mean and median values in the subsequent sections. \rm
In the online Table\,\ref{tab_period}, we list the rotation periods for each periodic target, together with uncertainties and normalised powers, determined in the individual time-series segments. \\  \rm
\indent
The light curves of all stars for which either the ASAS or the SuperWASP photometry allowed us to determine the rotation period are plotted in the online Figs.\,\ref{eps_cha_fig1}-\ref{ic2391_fig6}.\\
\indent

\begin{figure}
\begin{minipage}{10cm}
\centerline{
\includegraphics[trim=0 0 0 0,width=8.cm,height=8cm,angle=0]{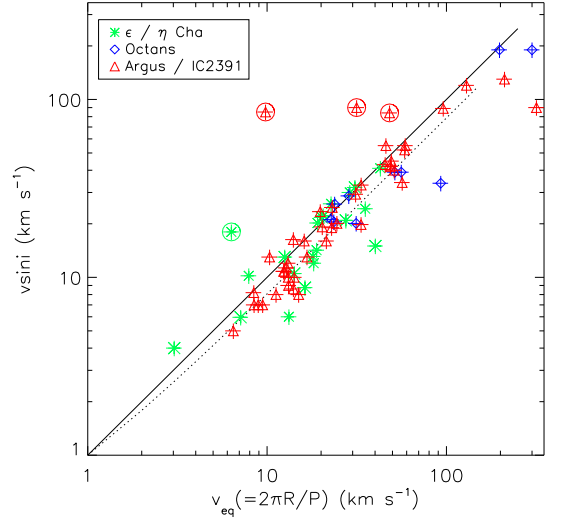}
}
\end{minipage}
\caption{$v\sin i$ from the literature vs. equatorial velocity  v$_{\rm eq}$ = 2$\pi$R/P. The solid line marks $v\sin i$ = v$_{\rm eq}$, whereas the dotted line  $v\sin i$ = ($\pi$/4)v$_{\rm eq}$. Circled symbols are stars whose  $v\sin i$ is inconsistent with  v$_{\rm eq}$\rm.\label{vsini} }
\end{figure}

\begin{figure*}[t]
\begin{minipage}{17cm}
\centerline{
\includegraphics[trim=0 0 0 0,width=7.5cm,height=8.5cm,angle=0]{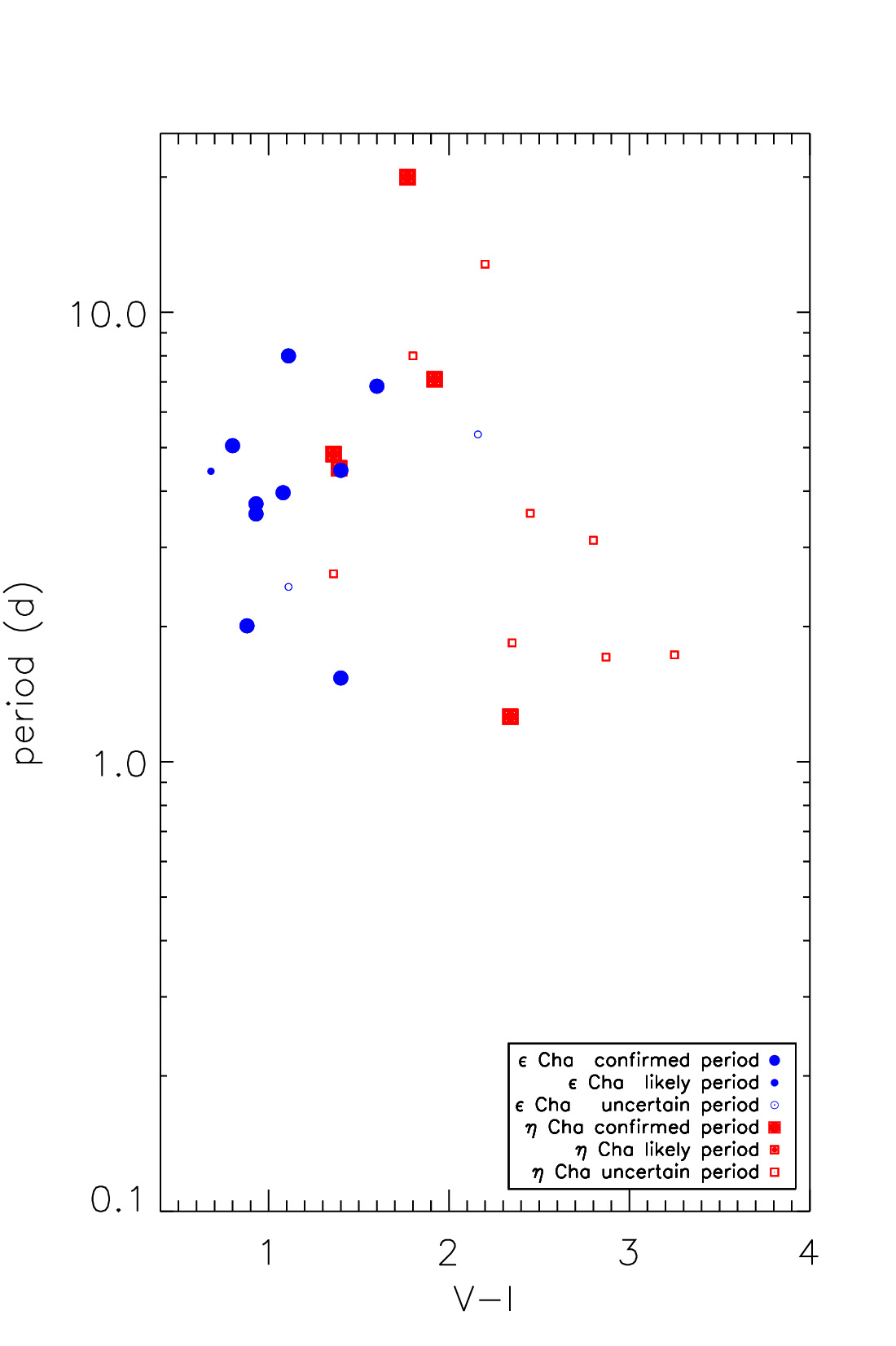}
\includegraphics[trim=0 0 0 0,width=7.5cm,height=8.5cm,angle=0]{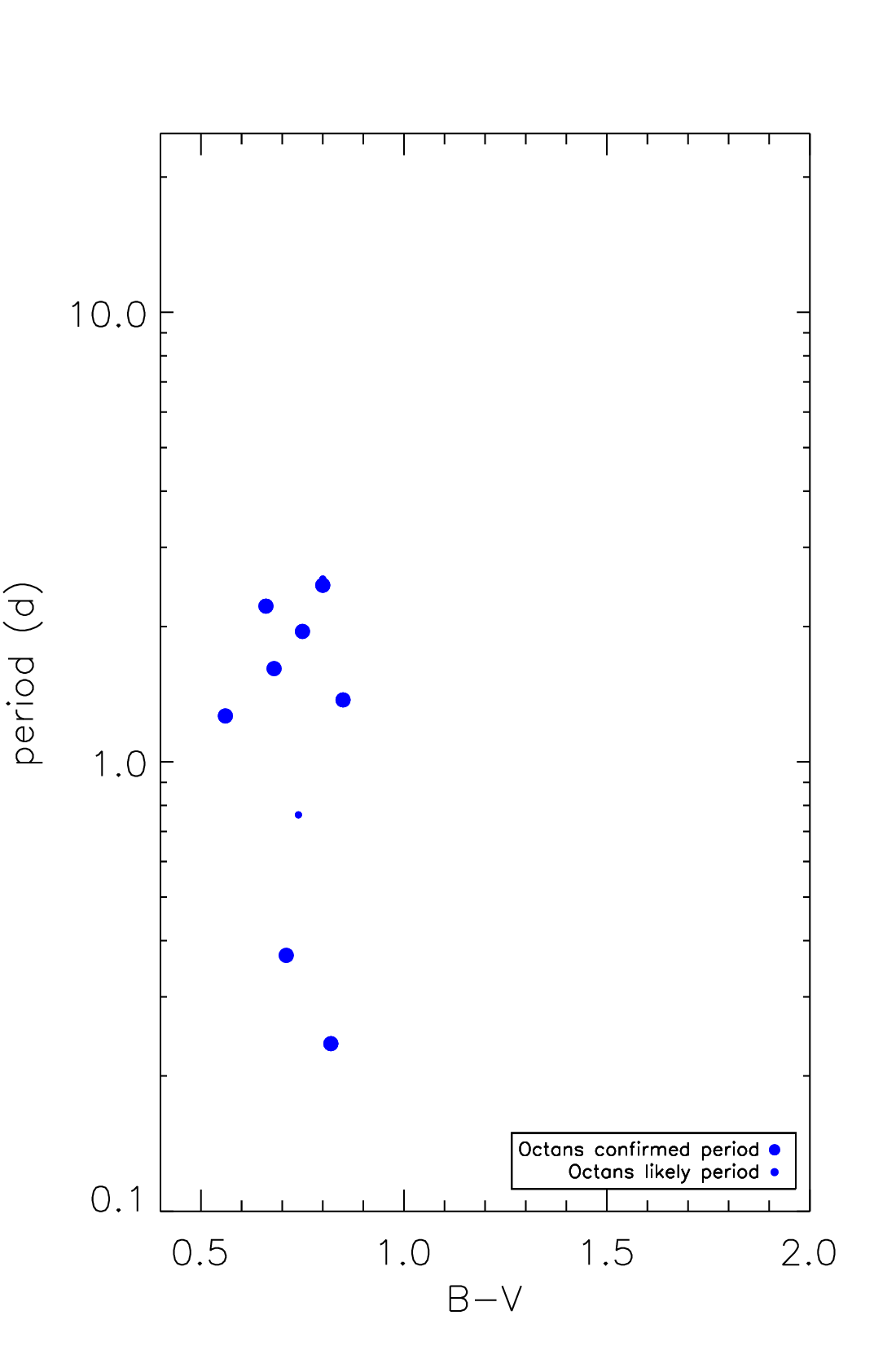}
}
\end{minipage}
\caption{\it left panel: \rm distribution of rotation period of $\epsilon$/$\eta$ Cha members versus V$-$I colour.  The V$-$I colour is used since almost all the $\eta$ Cha members lack B$-$V measurements.  \it Right panel: \rm same as left panel but for Octans members versus B$-$V. \label{distri}}
\end{figure*}

\subsection{Colour-magnitude diagrams}

Following the approach of Paper I, we use the M$_V$ vs. V$-$I CMD and a set of low-mass PMS evolutionary tracks 
to derive masses and radii. Owing to the targets distances, 
on average larger than 100 pc, and their young ages, especially  in the case of the $\epsilon$ / $\eta$ Cha members, 
the observed colours may suffer from reddening arising either from interstellar or circumstellar material. We have 
investigated on the possible colour excess of all targets and derived the intrinsic colours for the subsequent analysis.
V magnitudes are 
taken from ASAS because the long-term monitoring gives the possibility of measuring the brightest magnitude over 
a long time range,  i.e. the value corresponding to the observed minimum spot coverage. In most cases, the ASAS V-magnitudes 
are found to be  brighter than those reported in the references below.\\
\indent
{\it $\epsilon$ Chamaeleontis}.
Colours are taken either from Torres et al. (\cite{Torres06}) or Alcal\`a et al. (\cite{Alcala95}) both in the Johnson-Cousins photometric system. 
 All but two members (GSC\,9235-01702
and CD$-$74712)  are found to be affected by reddening (see online Tables\,\ref{cha_lit}-\ref{argus_lit}). 
The colour excess of each target is determined by comparing the observed V$-$I colour with the V$-$I colour corresponding 
to the spectral type of a standard dwarf star.  The A$_V$ extinction is derived from E(V$-$I) using the  relations of 
Mathis (\cite{Mathis90}).  We find that the  A$_V$ extinction ranges from 0.1 to 1.8 mag.
A study by Knude \& Hog (\cite{Knude98}) based on the Hipparcos and Tycho data found that in the Chamaeleon region 
the interstellar extinction  in the distance range  of our targets  never exceeds A$_V$$\sim$0.15 mag. Therefore,
 the derived reddening, in agreement with the very young age of the $\epsilon$ Cha members, likely arises from 
circumstellar rather than from interstellar  material.\\
\indent
{\it $\eta$ Chamaeleontis}.
Colours in the Johnson-Cousins photometric system and spectral types in the present paper are taken from Lawson et al. (\cite{Lawson01}; \cite{Lawson02}).
A study by Westin (\cite{Westin85}) has found the interstellar reddening for this cluster to be unimportant. This result is also 
confirmed by Mamajek et al. (\cite{Mamajek00}) who found the V$-$I colour of members to be consistent with the inferred spectral type.\\
\indent
{\it Octans}.
The observed colours are taken from Torres et al. (\cite{Torres06}) and are found to be 
consistent with the spectral types; therefore, no reddening correction is applied. We adopt the V$-$I colour
corresponding to the target's spectral type for four stars missing V$-$I measurements. Only one star (TYC\,7066\,1037\,1) with measured V$-$I colour shows an excess that 
we corrected as described above.\\
\indent
{\it Argus}. V$-$I colours are taken from Torres et al. (\cite{Torres06}).
We adopt the colours corresponding to the spectral type of
 standard dwarf stars for 14 members that lack V$-$I measurements. Only one star (TYC\,8561\,0970\,1) with measured V$-$I colour shows an excess that 
we corrected as above.\\
\indent
{\it IC2391}.
Colours in the Johnson-Cousins photometric system are taken either from Patten \& Simon (\cite{Patten96}) or from Platais et al. (\cite{Platais07}). In the latter case only B$-$V is measured and V$-$I is inferred from spectral type.  The mean reddening towards IC 2391 is estimated to be close to zero (Patten \& Simon \cite{Patten96})
and no correction was made for interstellar reddening. Only one star (PMM\,1759) needed to be corrected to make V$-$I consistent with spectral type.\\
\indent
\rm
In general,  as expected from variability arising from spots and/or faculae, \rm the V-magnitude variability amplitude in our targets never exceeds $\sim$0.3 mag. The only deviating star is  PMM\,8145 whose variability amplitude reaches $\sim$2 mag. In this  case, however, there is a close, very bright nearby star ($o$ Vel, V=3.6 mag $\Delta \alpha$=2 arcsec, $\Delta \delta$=70 arcsec) and therefore CCD saturation affecting the PMM\,8415 time-series cannot be ruled out.\\
\indent
 V magnitudes, B$-$V, and U$-$B colours  (flagged with A$_V$ when corrected for reddening), \rm absolute magnitudes, distances and $v\sin i$ values from the literature are listed in the online Tables\,\ref{cha_lit}-\ref{argus_lit}.\\
\indent
The  reddening-corrected \rm CMDs of the studied associations are plotted in Figs.\,\ref{cmd}-\ref{cmd1} together with the Baraffe et al. (\cite{Baraffe98})
 evolutionary tracks and the CMDs of three well-studied open clusters of known age for reference: $\alpha$ Persei ($\sim$70 Myr), the Pleiades ($\sim$110 Myr), and NGC\,2516 ($\sim$150 Myr).\\
\indent
As in Paper I, we estimate masses and radii comparing positions in the CMD with the Baraffe et al. (\cite{Baraffe98}) evolutionary tracks.
The stellar radius (R) allows us to make a comparison between $v\sin i$ and the equatorial velocity v$_{eq}$=2$\pi$ R /P (where P is the rotation period)
to check the consistency between the two and derive the stellar inclination. Derived masses and radii are listed in the online Tables\,\ref{cha_lit}-\ref{argus_lit}.

\begin{figure}
\begin{minipage}{10cm}
\centerline{
\includegraphics[trim=0 0 0 0,width=7.5cm,height=8.5cm,angle=0]{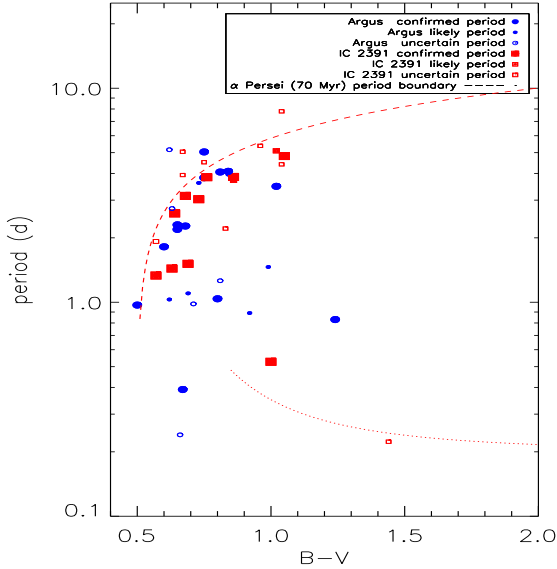}
}
\end{minipage}
\caption{The same as Fig.\,\ref{distri}, but for Argus / IC\,2391 association.\label{distri1} The dashed and dotted lines represent  mathematical functions  describing the loci of the upper and lower  bounds of the period distribution of $\alpha$ Persei members according to Barnes (\cite{Barnes03}).}
\end{figure}

\subsection{$v\sin i$ vs. equatorial velocity}
All but five of the periodic variables in our sample have known projected
equatorial velocities ($v\sin i$). 
In Fig.\,\ref{vsini}, we compare $v\sin i$ and v$_{eq}$, by delineating the loci of $v\sin i$=v$_{eq}$,
corresponding to equator-on orientation,  and  $v\sin i$=$\pi$/4 v$_{eq}$,
corresponding to a randomly orientated rotation axis distribution. In Fig.\,\ref{vsini} all periodic (\it confirmed\rm, \it likely\rm, and \it uncertain\rm) targets are plotted. The major
uncertainty in the equatorial velocity is the radius estimate. The
reported $v\sin i$ uncertainties are  on average 10\%. Only four stars ( GSC\,9419-01065 in $\epsilon$ Cha, \rm CD-582194 and CPD-621197 in Argus, and PMM\,351 in IC\,2391)
(flagged  with an apex  $c$ in Tables\,\ref{t:eps Cha} and \ref{t:argus} and plotted with circled symbols in Fig.\,\ref{vsini})
have inconsistent $v\sin i$/v$_{eq}$ (i.e., much larger than unity). These cases are discussed in Appendix A.

In Table\,\ref{tab-vsini}, for each association we report the average $<$sin$i$$>$$\pm$$\sigma$,  
the correlation coefficient $r$ from the linear Pearson statistics between $v\sin i$ and  v$_{eq}$, where the labels a and b indicate the 
significance level of the correlation coefficient. The significance level represents the probability of 
observing a value of the correlation coefficient greater than $r$ for a random sample with the same number of observations and degrees of freedom.\\
It is interesting to note that all association members have mean inclinations
that are inconsistent with the value ($\pi$/4) expected for a completely randomly  orientated
rotational axis distribution. Specifically, all association members tend to appear almost equator-on to
the observer. This behaviour, which was also exhibited by the  six other  associations within 100 pc analysed in Paper I, may arise
from some bias rather than being real.
A bias in the member selection is unlikely, since the association members were identified in
 larger samples selected on the basis of their X-ray emission, which is not affected by the inclination 
of the rotation axis. On the contrary, a bias towards equator-on members  may be more likely.
In fact,  we find that members with unknown rotation periods exhibit on average $v\sin i$ systematically
smaller than members with \it uncertain, likely, \rm and \it confirmed \rm  periods, respectively. In the specific 
case of Argus / IC 2391, which is the most numerous association in the present paper,
we find that members with unknown periods have $<$$v\sin i$$>$=15.1 km s$^{-1}$ with respect to $<$$v\sin i$$>$=33.8 km s$^{-1}$
exhibited by members with \it confirmed  \rm periods. In the case of the AB Dor association studied in Paper I we find
$<$$v\sin i$$>$=8.2 km s$^{-1}$ for stars with unknown period against $<$$v\sin i$$>$=25.9 km s$^{-1}$
for members with \it confirmed  \rm periods. From the former case, we can infer that about  20\% of stars, whose average $v\sin i$ is about half 
 that of equator-on stars, is still missing in the $v\sin i$ versus v$_{eq}$ plot of Argus / IC 2391.
Therefore, a bias towards stars with high values of inclination is present, which are those 
that show larger-amplitude light modulation and that are most favoured for the rotation period determination.
We may also suppose that our derived radii are all systematically underestimated,
 likely from  effects of high magnetic activity and fast rotation, as found
 by, e.g.,  Chabrier et al. (\cite{Chabrier07}) and Morales et al. (\cite{Morales09}). This makes v$_{\rm eq}$ systematically small 
with respect to the measured $v\sin i$. Nonetheless,
these two quantities are found to be strongly correlated (see Table \ref{tab-vsini}) differently than expected in the case of random distribution
of axes. This finding certainly deserves further investigation to take all possible observation biases  into account
so is beyond the scope of the present paper.\rm

\subsection{$\epsilon$ / $\eta$ Chamaeleontis}
We determined the rotation periods for 11 stars in $\epsilon$ Cha, of which there are ten new and one already known in the literature. The rotation periods of another  two stars were retrieved from the literature. For stars in $\eta$ Cha, we retrieved 12 rotation periods, of which seven were retrieved from the literature. For $\epsilon$ and $\eta$ Cha we  therefore have a total of 25 periods available, 14 \it confirmed, \rm  1 \it likely, \rm and 10 \it uncertain. \rm
In the left panel of Fig.\,\ref{distri}, we plot with filled blue bullets and red squares the  members of $\epsilon$ and $\eta$ Cha, respectively,  with \it confirmed \rm periods. Small-size filled symbols represent \it likely \rm periods, whereas small-size open symbols are used for \it uncertain \rm periods.\\
\indent
Members of $\epsilon$ Cha with \it confirmed \rm periods have   all  M $\ge$ 0.8M$_{\odot}$. \rm Their periods have a mean P$_{\rm mean}$ = 4.35d and a median P$_{\rm median}$ = 3.97d. Members of $\eta$ Cha with confirmed periods have all but one M$\ge$0.8M$_{\odot}$. Their periods have P$_{\rm mean}$ = 4.67d and P$_{\rm median}$ = 4.84d, although only derived from four stars. If $\epsilon$ and $\eta$ Cha are considered together as a coeval system, we obtain P$_{\rm mean}$ = P$_{\rm median}$ = 4.45d.

\subsection{Octans}
Of 12 late-type members to the Octans association, we determine the rotation periods of ten stars (eight of which are \it confirmed \rm
periods, and two are \it likely \rm periods). The Octans rotation period distribution is plotted in the right panel of Fig.\,\ref{distri}.
Owing to the very small number of known late-type members, the period distribution upper bound is not sufficiently defined. 
All \it confirmed \rm periodic members in our study have masses M $\ge$ 0.8M$_{\odot}$ and the mean and median periods turned out to be P$_{\rm mean}$ = 1.46d and P$_{\rm median}$ = 1.61d.\\
\indent
 Octans is the most distant association ($\sim$140 pc) and, owing to the small number of members with accurate trigonometric parallax,
 its age is the most uncertain in our sample. As for the other associations, the Octans members were selected by SACY on the basis of their X-ray emission 
in the ROSAT Bright Survey. However, the detected members of Octans are at the detection limit of that survey. Therefore,
 it is likely that only the most active and, therefore, fast
rotating members have been detected. That means that our sample of periodic stars is biased, the detection of only the fastest rotating stars being favoured.
Therefore, the bias towards the faster rotators, together with the uncertainty on age, makes the Octans mean and median rotational periods quite unreliable, which must be taken into account in our analysis of the rotation period evolution reported in Fig.\,\ref{age}. \rm

\subsection{Argus / IC\,2391}
Among the three associations under analysis, Argus has the largest number of late-type members, 27 Argus stars plus 30 additional stars in the IC\,2391 cluster. We  determine
the rotation period of 45 members. However, only for 23 members, which are plotted in Fig.\,\ref{distri1}, are the rotation periods   \it confirmed\rm. \\
\indent
 We overplot the empirically determined functions computed according to Barnes (2003) and describing the loci of the upper/lower bounds of the $\alpha$ Persei period distribution for main-sequence stars. \rm
The Argus/IC\,2391 upper bound is very close to the dashed line, which is computed for a nominal age of 70 Myr, suggesting an age for Argus/IC\,2391, slightly older than the quoted age of 40 Myr.
Considering \it confirmed \rm periods only, we found P$_{\rm mean}$ = P$_{\rm median}$ = 2.27d for Argus, P$_{\rm mean}$ = 2.36d, and P$_{\rm median}$ = 3.03d for IC\,2391. 
In both cases all stars have mass M $>$ 0.8M$_{\odot}$. When all members (Argus/IC\,2391) are considered, together P$_{\rm mean}$ = 2.43 and P$_{\rm median}$ = 2.30d.
 Although quoted with the same age, we see,  either in Fig.\,\ref{distri1} or from the above mean values, that the distribution of rotation periods of IC 2391 
stars appears slightly different from that of Argus stars, with more slow rotators in the cluster. This difference may arise from the
different selection criteria adopted to search for members. Argus members are identified on the basis of their X-ray emission and are all present in the
ROSAT Bright Source Catalog (Voges et al. \cite{Voges99}). Conversely, the IC 2391 members  are selected from a variety
of sources, without X-ray preselection. Since fast rotators have brighter coronal luminosities and the mean distance of Argus asociation members is 
106 pc, compared to $\sim$145 pc for IC\,2391, we expect that the census of association members is biased toward the most active and fast-rotating stars (see Desidera et al. \cite{Desidera11}). \rm

\begin{figure*}[t,*h]
\begin{minipage}{18cm}
\centerline{
\includegraphics[trim=0 0 0 0,width=6.5cm,height=6.5cm,angle=0]{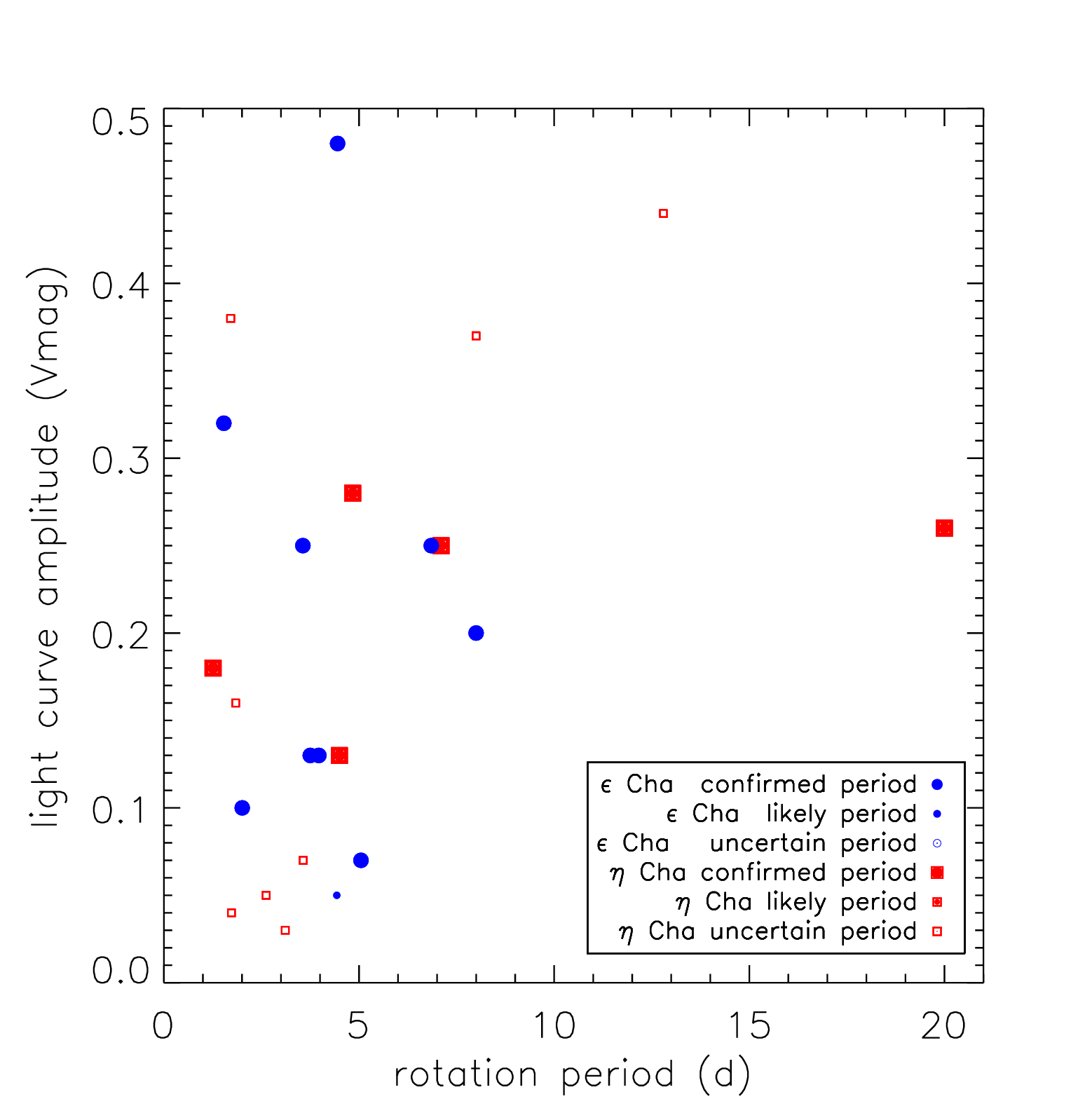}
\includegraphics[trim=0 0 0 0,width=6.5cm,height=6.5cm,angle=0]{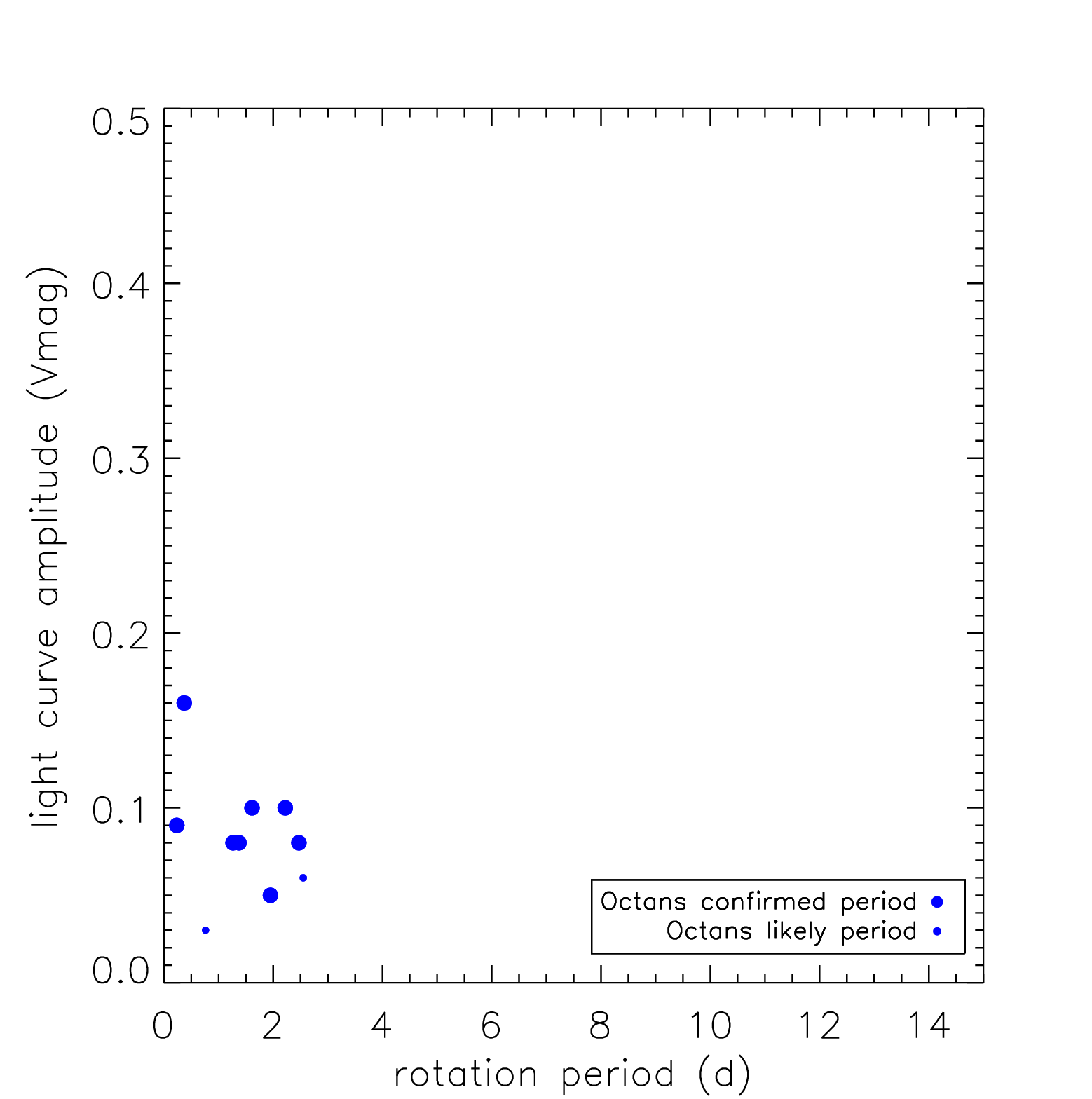}
\includegraphics[trim=0 0 0 0,width=6.5cm,height=6.5cm,angle=0]{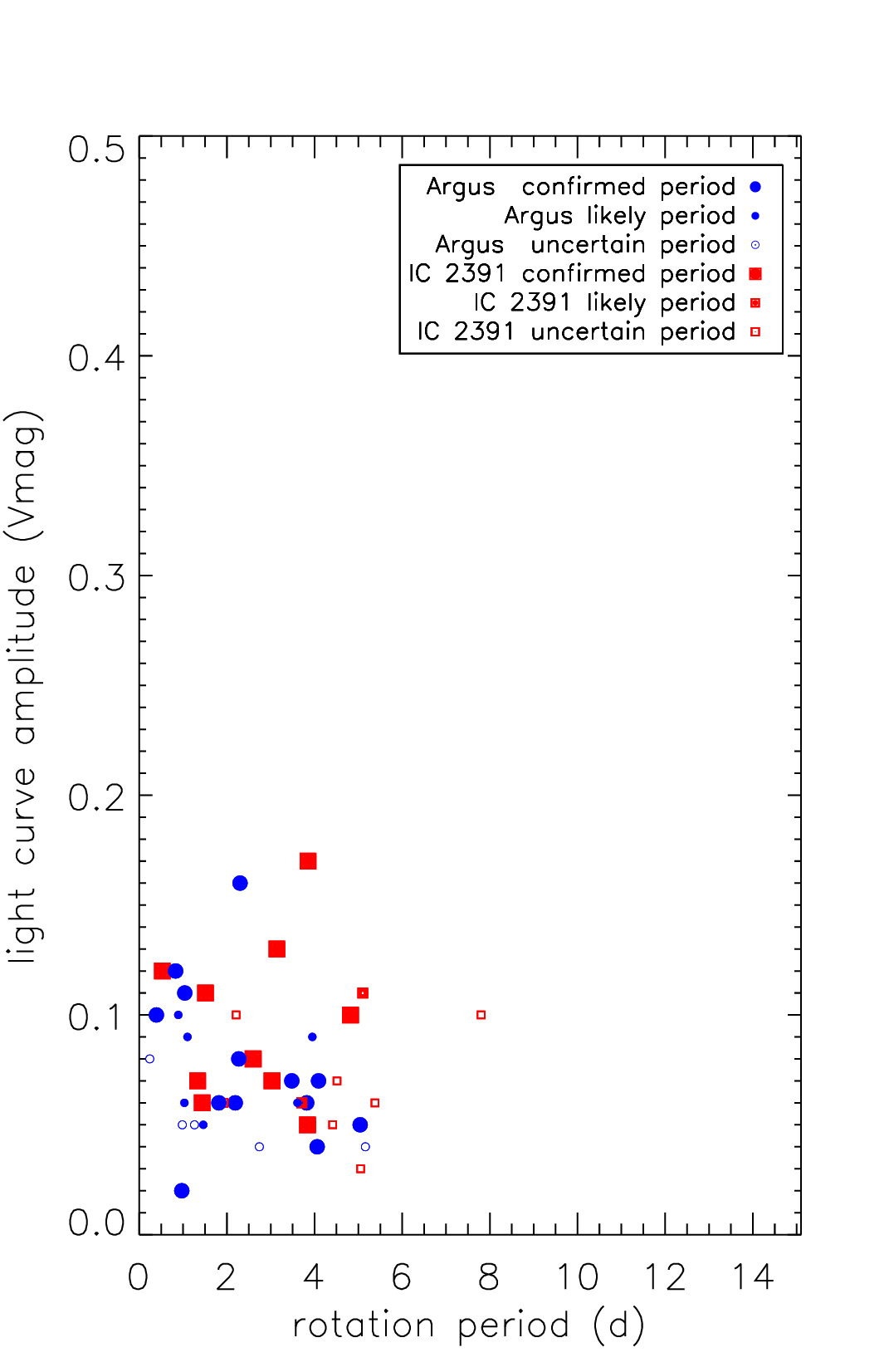}
}
\end{minipage}
\caption{Distribution of V-band peak-to-peak light curve amplitudes versus rotation period for the members of $\epsilon$/$\eta$ Cha (left panel), Octans (middle panel), and Argus/IC\,2391 (right panel.\label{amp_new})}
\end{figure*}

\subsection{Rotation-photospheric activity connection}
Following Paper I, we have extracted for each \it confirmed \rm  periodic target the V-band maximum peak-to-peak light curve amplitude ever observed in the available time series  that are plotted in Fig.\,\ref{amp_new}.
For the older associations under analysis, Octans and Argus/IC\,2391, the photometric variability mostly arises from the presence of cool spots
on the stellar photosphere. In the case of $\epsilon$/$\eta$ Cha, a contribution from hot spots, arising from accretion processes, cannot be
ruled out. The largest amplitudes are observed in the youngest association  at an age of about 6 Myr where  the presence of 
circumstellar material (as inferred from the observed colour excess) and related accretion phenomena are still expected. The lowest amplitudes are observed in the Octans association
with an assumed age of about 20 Myr. The apparent age dependence of the light curve amplitude at fixed rotation and mass will be discussed in a forthcoming
paper (Messina et al., in preparation).

\subsection{Rotation period evolution}
 
In Fig.\,\ref{age}, we plot the results of our period search in the nine associations under study, where we show the rotation period variation of low-mass stars versus age in the 0.8-1.2 M$_{\odot}$ range. We plot the rotation periods listed in Tables 2-4 of this paper, and in Tables 3-8 of Paper I, complemented with the updated values listed in Table B.1 of this paper. We overplot the median and mean rotation periods computed for each association (or coeval associations as explained by the labels) considering only the \it confirmed \rm
rotation periods to which we assign the same weight. We also add the mean and median rotation periods of ONC (Herbst et al. 2002), NGC\,2264 (Rebull et al. 2002; Lamm et al. 2004),
$\alpha$ Persei, and the Pleiades, whose rotation periods are taken from the compilation of Messina et al. (2003, and references therein).\\
\indent
In Table \ref{tab-per} we report for each cluster/association the number of confirmed periods used to derive the mean and median values, as well as the KS significance levels that consecutive (at increasing ages) measured period distributions are drawn from the same distribution according to Kolmogorov-Smirnov tests (see, e.g., sect. 14.3 of Press et al. (\cite{Press92}). Period distribution variations that are not statistically significant have KS values close to unity, while statistically significant variations have KS values close to zero.

The newly added mean and median values at 6 Myr ($\epsilon$̨/$\eta$ Cha) and 40 Myr (Argus/IC 2391) confirm the trend of period evolution with age found in Paper I. From 1 to 9 Myr, the mean and median periods slowly decrease, but the KS values remain rather high, which indicate that indeed the period distributions do not change significantly. This behaviour is consistent with a locking mechanism operating in this age range. From 9 to 30 Myr both mean and median periods decrease and the KS values indicate a statistical significant variation in this case. The decrease in mean and median periods is monotonic, with only the period distribution of Octans deviating from this trend. However, as discussed in Sect. 5.4, the sample of known Octans late-type components is expected to be rather incomplete and biased towards fast rotators, which could explain the rather low mean and median periods compared with adjacent values. Furthermore, the Octans age uncertainty is considerably larger than for the other associations. For the 0.8 - 1.2 M$_{\odot}$ range, our analysis is therefore consistent with a considerable disc-locking before 9 Myr, followed by a moderate but unambiguous spin-up from 9 to 30 Myr, consistent with stellar contraction towards the ZAMS. 

Variations between 30 and 70 Myr are rather doubtful. The KS test indicates that there is no significant variation between 30 and 40 Myr.  On the other hand, the KS test indicates a significant variation between 40 and 70 Myr, but while the median indicates a significant spin-up, the mean remains approximately constant. Between 70 and 110 Myr the KS test indicates a significant variation and both mean and median periods increase.  This situation may be due to the heterogeneity of the sample: all stars with masses above 1 M$_{\odot}$ are expected to complete their contraction toward the ZAMS at ages earlier than about 30 Myr, but stars with lower mass will end the contraction towards the ZAMS later on (around 70 Myr for a star of 0.8 M$_{\odot}$). The unambiguous spin-down from 70 to 110 Myr is consistent with all stars in the 0.8 - 1.2 M$_{\odot}$ mass range having entered the MS phase and therefore the angular momentum evolution being dominated by wind-braking.

\rm

\section{Conclusions}
We have performed a rotation period search for all late-type members of the nine young ($<$100 Myr) associations known to date for which either ASAS or SuperWASP data are available. We supplemented such information with rotation periods retrieved from the literature to derive a catalogue that contains (considering also the new periods listed in Table B.1) a total of 241 rotation periods. We have established quality criteria for classifying the derived period based on the frequency of period determination in various light-curve segments, independent measurements retrieved from the literature, and consistency with $v\sin i$. Based on such criteria, three quality levels are proposed: confirmed, likely, and uncertain. Our catalogue contains 171 \it confirmed\rm, 44 \it likely\rm, and 26 \it uncertain \rm periods. The rotation period remains unknown for the remaining 50 late-type members.  Thanks to the newly determined period distributions at $\sim$6, $\sim$30, and $\sim$40 Myr, our catalogue allows a better empirical description of angular momentum evolution of stars with masses from 0.8 to 1.2 M$_{\odot}$ and with ages from 1 to 100 Myr. \rm

The catalogue is used to build rotation period distributions vs. colours for each association in the 0.8-1.0 M$_{\odot}$ mass range. Excluding Octans, which is likely to be affected by a strong selection bias toward shorter periods, the average and median periods are found to essentially decrease from the ONC age ($\sim$1 Myr) untill the $\alpha$  Persei age ($\sim$70 Myr). The two-sided KS test indicates that indeed the period distributions do not change much from 1 to 9 Myr, while variations are statistically significant from 9 to 30 Myr. This increase in the average rotation rate with age is consistent with the contraction of the stars toward the ZAMS, which is contrasted by disc-locking at an early stage. The situation from 30 to 70 Myr is rather uncertain, probably because of the heterogeneity of our sample, in which stars of different mass reach the ZAMS at different ages. From $\alpha$ Per (70 Myr) to the AB Dor/Pleiades (110 Myr) both mean and median periods increase with the KS test, indicating a statistically significant variation.

\begin{table*}
\caption{ Summary of rotation period average values and of Kolmogorov-Smirnov test (KS) results.\label{tab-per}}
\begin{tabular}{lc|cccc|cccc|l}
\hline
					&	& \multicolumn{4}{c|}{\bf  0.6-1.2 M$_{\odot}$} & \multicolumn{4}{c|}{\bf 0.8-1.2 M$_{\odot}$} &  \\
\hline
Target    & Age  &  \#     &     P$_{\rm median}$     &     P$_{\rm mean}$ &  KS  & \#     &     P$_{\rm median}$     &     P$_{\rm mean}$ &  KS  &  \\
              & (Myr) & Stars  &	(d)					& (d)			& & Stars  &	(d)					& (d)			&\\
\hline
ONC       					&  1  	&   33   &   6.83   &    6.17 & 0.57 &  16    & 6.83   & 5.93 & 0.70  & 1 vs. 4 Myr\\
NGC\,2264     				&  4    	&   26   &   5.43   &    6.19 & 0.59 &  10    & 4.72   & 5.15 & 0.63  & 4 vs. 6 Myr\\
$\epsilon$ / $\eta$ Cha		&  6		&   ...    &	...		& ...	& ...	&       14	   & 3.97	&	4.35 	& 0.30&  6 vs. 9 Myr\\
TW Hya / $\beta$ Pic    	        &  9  	&   37   &   4.83   &    4.75 & 0.09 &  21    & 4.27   & 4.43 & 0.02  & 9 vs. 20 Myr\\
Octans					& 20		&   ...    &	...	    & ...	   & ...    &    8	   & 1.61   & 1.46 & 0.12     &  20 vs. 30 Myr\\	
Tuc/Hor / Car / Col     		&  30 	&   53   &   2.72   &    2.88 & 0.36 &  45    & 2.57   & 2.69 & 0.96  & 30 vs. 40 Myr\\
Argus / IC\,2391			& 40		&  ...    &	...	    & ...	   & ...    &   23	&  2.30    & 2.43  & 0.00 & 40 vs. 70 Myr\\
$\alpha$ Persei    			&  70  	&   54   &   0.77   &    2.39 & 0.12 &  48    & 0.77   & 2.49 & 0.11  &70 vs. 110 Myr\\
AB\,Dor / Pleiades    		&  110 	&   75   &   1.49   &    2.73 & ...     & 49    & 1.79   & 2.99 & ...     & ... \\
\hline

\end{tabular}
\end{table*}

\begin{figure}[]
\begin{minipage}{10cm}
\centerline{
\includegraphics[trim=0 0 0 0,width=9cm,height=8.5cm,angle=0]{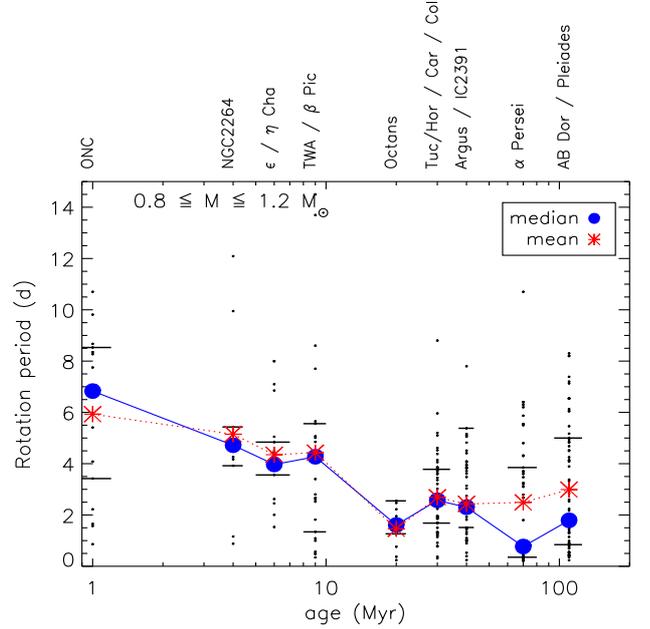}
}
\end{minipage}
\caption{Rotation period evolution versus time in the 0.8-1.2 solar mass range. Small dots represent individual rotation period measurements. Bullets connected by
solid lines are median periods, whereas asterisks connected by dotted lines are mean periods. Short horizontal lines represent the 25th and 75th percentiles 
of rotation period. This plot updates the right panel of Fig.\,12 of Paper I.  \label{age}}
\end{figure}

\section*{Acknowledgements}
The extensive use of the SIMBAD and ADS  databases  operated by  the  CDS centre,  Strasbourg,
France,  is gratefully  acknowledged. 
We  used data from the WASP public archive in this research. The WASP consortium comprises of the 
University of Cambridge, Keele University, University of Leicester, The Open University, The Queen's University Belfast, 
St. Andrews University, and the Isaac Newton Group. Funding for WASP comes from the consortium universities and 
from the UK's Science and Technology Facilities Council.
 The Authors would like to thank Dr.\,G. Pojma\'nski for the extensive use
we made of the ASAS database. We are grateful to the Referee Dr. James for his valuable comments that
allowed us to significantly improve our analysis and its presentation.

\appendix

\section{Individual stars}

\subsection{$\epsilon$ + $\eta$ Cha}

\indent 

\bf GSC\,9419-01065: \rm the rotation period derived from our analysis is P=8.0d. Although
the flux rotational modulation is clearly found in more than five segments; however, the computed  v$_{\rm eq}$=2$\pi$R/P
is inconsistent with the measured $v \sin i$=18.0 km s$^{-1}$ (Torres et al. \cite{Torres06}), so  this period 
is considered \it uncertain\rm. However, we note  that the only available $v \sin i$ measurement 
has  quite a large $\pm$30\% uncertainty. More accurate determinations would be desiderable
to confirm the correctness of the current $v \sin i$ value.


\bf HD\,104237E: \rm the ASAS photometry did not allowed us to infer any periodicity. 
We adopted the period of P=2.45d detected by Feigelson et al. (\cite{Feigelson03}), which combined
with the estimated radius gives an equatorial velocity consistent with the measured $v\sin i$.
This star belongs to a quintuple system, where the brightest component is the Herbig Ae star HD\,104237.
Feigelson et al. (\cite{Feigelson03}) found a reddening A$_{\rm V}$=1.8$\pm$0.3 that likely
arises from material within the parent stellar system, the reddening of other members associated to $\epsilon$ Cha
being small or even absent.


\bf GSC\,9416-1029: \rm Doppmann et al. (\cite{Doppmann07}) report a period of P=5.35d that likely represents the system's orbital period. 
The available ASAS photometry of this star, which is quite inaccurate owing to the target's  faintness, shows evidence of a 
period P=5.50d in two of nine time segments. In the case this is the correct rotational period, which would imply a rotational/orbital synchronisation,
which is  quite a surprising circumstance at an age of about 6 Myr.

\bf GSC\,9235-01702: \rm it has very recently been discovered to be a  member of $\epsilon$ Cha by Kiss et al. (\cite{Kiss11}) as part of the RAVE project. 
From the ASAS photometry, we found the same rotation period as found by Bernhard et al. (\cite{Bernhard09}), based on the same ASAS data.

\bf RECX\,1 (EG Cha): \rm we detect a rotation period P=4.5d in 12 of 15 time segments, which is twice the period reported by Lawson et al. (\cite{Lawson01}).
An inspection of Fig.\,2 of Lawson et al. (\cite{Lawson01}) shows that their phased light curves  in both 1999 and 2000 seasons exhibit
significant magnitude phase dispersion.

\bf RECX\,11 (EP Cha): \rm we found a period P=4.84d in 7 of 12 time segments with confidence level over 99\% and no evidence of any power peak
at the period P=3.95d  discovered by  Lawson et al. (\cite{Lawson01}) in 1999 (P=3.69d in 2000). Again, an inspection of their phased 
light curves shows significant dispersion

\bf RECX\,12 (EQ Cha): \rm the star shows two significant periods in Lawson et al.~(\cite{Lawson01}), P=1.25d and P=8.55d. The star is a close  binary,
so they might represent the rotational periods of the two components. Our analysis allowed us to detect only the shorter period that is adopted in the
present analysis.

\bf RECX\,15 (ET Cha): \rm is a classical T Tauri star, with evidence of on going accretion (Lawson et al. \cite{Lawson02}). The large amplitude variations
(up to 0.44 mag) are likely driven by accretion hot spots.

\subsection{Octans}

\indent 

\bf CD-58\,860: \rm we find two periods, P=1.612d and P=2.610d, in eight out of ten time segments with comparable levels of
confidence. However, the longer period, once combined with stellar radius, determines a ratio $v\sin i$/v$_{\rm eq}$
slightly greater  than unity. Therefore, in the present analysis we  adopt P=1.612d.

\bf CD-43\,1451: \rm this star is also present in the SuperWASP archive. However, 
neither from the ASAS nor from the SuperWASP database we could determine the rotation period,
but only the presence of light variability.

\bf HD\,274576: \rm the period P=2.22d is found in nine out of ten time segments of ASAS photometry and in two out of two SuperWASP observation seasons,
and it is fully consistent with the measured $v\sin i$. We note that in both the ASAS and SuperWASP timeseries another period
P=1.824d is also found with a very high confidence level, although it is smaller than the earlier,
in almost all time segments, which is also consistent with $v\sin i$. 

\bf TYC\,7066\,1037\,1: \rm the period P=2.47d is found in 6 out of 14 time segments of ASAS photometry and in three out  of five
segments of  SuperWASP photometry.

\indent

\subsection{Argus and IC\,2391}

\indent 

\bf HD\,5578 (BW Phe): \rm is a close visual binary. We found two rotation periods of comparable power, P=1.461d and P=3.15d, in most time segments.
The shorter period is assumed to be the star's rotation period, since it is  the only one consistent with a $v\sin i$/v$_{\rm eq}$$\le$1.


\bf CD-56\,1438: \rm   we detect a period P=0.24d in only one segment that may conciliate with the very high $v\sin i$.

\bf CD-28\,3434: \rm  the P=3.82d period is very well established and also detected in three out  of three time segments of SuperWASP data, as well in
the complete timeseries. Although no $v\sin i$ is available to check consistency with v$_{\rm eq}$, it is considered a  \it confirmed \rm period.

\bf HD\,61005: \rm has been suggested to be a likely member of the Argus association (Desidera et al. \cite{Desidera11}). Although its P=5.04d period is found
in only three time segments of the ASAS timeseries, it was confirmed by the period search carried out in the Tycho and Hipparcos photometry (Desidera et al. \cite{Desidera11}), 
so it  is considered \it confirmed \rm in the present analysis. \rm

\bf CD-39\,5883: \rm  the period is detected in two out  of nine segments of  ASAS photometry and in five out of five segments of WASP data
as well in the complete timeseries and is consistent with $v\sin i$/v$_{\rm eq}$$\le$1.


\bf CD-58\,2194: \rm  the most significant period is P=5.16 detected in four time segments as well in the complete timeseries.
However, it is inconsistent 
with the high $v\sin i$. It may be the beat of the P=0.55d period 
detected in only one season. This period is therefore considered \it uncertain\rm.

\bf CD-57\,2315: \rm  it is variable, but no periodicity was found.

\bf CPD-62\,1197: \rm  the most significant period is P=1.26d which is found in seven out  of nine segments. However, it leads to a $v\sin i$/v$_{\rm eq}$$\sim$2
and therefore it will not be considered in the following analysis. Another significant period is P=0.82d, which is, however only detected  in three out of nine segments,
so it is classified as \it uncertain.\rm

\bf TYC\,7695\,0335\,1: \rm  the same P=0.39d is found in three out  of nine ASAS segments and in four out of four segments of SuperWASP data.


\bf HD85151A: \rm  is a close visual binary. The P=0.97d is only found in the ASAS complete timeseries, but in five out  of five  time segments of SuperWASP data.

\bf CD-65\,817: \rm is a close visual binary.

\bf HD 310316: \rm is a close visual binary.

\bf CD-74\,673: \rm is a spectroscopic binary with an orbital period P=614d (Guenther et al. \cite{Guenther07}).

\bf CD-52\,9381: \rm  the most significant period is P=5.19d detected in six out  of nine  time segments, as well in the complete series.
However,  a P=0.89d is also  detected and reported in the ACVS. The shorter one is consistent with $v\sin i$/v$_{\rm eq}$$\le$1,
and it is  considered as a \it confirmed \rm period.

\bf PPM\,351: \rm the most significant period is P=1.931d that is found in six out  of nine segments; however, it gives $v\sin i$/v$_{\rm eq}$$>$1. Another detected period is P=0.69d, which is, however, found in only three out  of nine segments, so it is  considered \it uncertain. \rm

\bf PMM\,1083 (V365 Vel): \rm our period determination is in good agreement with the earlier determination by Patten \& Simon (\cite{Patten96}). Two different $v\sin i$ values are reported in the literature, $v\sin i$=43 kms$^{-1}$ from Marsden et al. (\cite{Marsden09}) and $v\sin i$=67 kms$^{-1}$ from Platais et al.~(\cite{Platais07}). However, only the first is consistent with $v\sin i$/v$_{\rm eq}$$\le$1.

\bf PMM\,1820 (V366 Vel): \rm our period determination is in good agreement with the earlier determination by Patten \& Simon (\cite{Patten96}).

\bf PMM\,4413: \rm is an SB2 with an orbital period P=90.6d (Platais et al.~\cite{Platais07}) whose components have measured $v\sin i$  of 8.6 and 8.4 kms$^{-1}$, which give 
consistent $v\sin i$/v$_{\rm eq}$ ratios.

\bf PMM\,4467 (V364 Vel);  PMM\,4902; PMM\,5884 (V377 Vel): \rm our period determinations are in good agreement with the earlier determinations by Patten \& Simon (\cite{Patten96}).

\bf PPM\,8145: \rm unlike stars whose variability arises from dark spots, this star spends most of its time in its
fainter state. The variability likely arises from magnitude outbursts. It shows the largest variability
amplitude ($\sim$ 2 mag) in our sample. The reference magnitude, differently than other stars, is probably the faintest
one.

\bf PMM\,4902 (V379 Vel): \rm  although detected in only three  time segments, it is considered \it confirmed \rm  because its period is confirmed by the 
literature value (Patten \& Simon \cite{Patten96}).

\bf PPM\,2182: \rm  we found two significant periods, P=3.28d and P=1.437d. However, both are not  consistent with  
$v\sin i$/v$_{\rm eq}$$\le$1, being $v\sin i$=78 kms$^{-1}$ (da Silva et al. \cite{daSilva09}) and therefore are considered 
\it uncertain. \rm

\section{New/revised periodicities in young associations studied in Paper I}

The availability of SuperWASP light curves allowed us to revisit some of the targets studied in Paper I.
We found SuperWASP timeseries for 71 targets listed in Paper I.
For the seven candidate members of $\beta$ Pic and Tuc/Hor newly proposed by Kiss et al. (\cite{Kiss11}), we retrieved SuperWASP timeseries
for two of them and ASAS time series for the remaining five.
Unfortunately, of 78 targets  the data on nine stars turned out to be too sparse to be suitable for a meaningful period search. The analysis of the 69 timeseries allowed us to discover (\it i\rm) 15 new periods (5 of which \it likely\rm), (\it ii\rm) to confirm 35 periods, and (\it iii\rm) to revise 13 periods. Finally, for six stars we could not detect any period. 
Three of them were also found non periodic in Paper I, two were found periodic in only 1 or 2 ASAS time segments, 
and one had only one literature determination. Our results are summarised in Table\,\ref{period-revisited}.

Concerning the revised periods, the previous determinations reported in Paper I were either  based on literature values (1 target),
had inconsistent $v\sin i$/v$_{\rm eq}$ ratio (4 targets), or were beat periods detected in less than five time segments (5 targets), or detected in five or more segments (3 targets).
Based on this result for the period revision of a few targets in Paper I, in the present work we decided to consider a rotation period to be well established (\it confirmed\rm) if detected
in five or more time segments or in fewer, but with an independent determination from the literature.

In the following we update the results presented in Paper I including new results based on Super WASP and on a more conservative period selection. \\

\indent
{\bf TW Hydrae}\\
We found SuperWASP data for eight targets in TW Hya and confirmed the rotation periods of four targets,
and determined three new periods. However, only one out of the three newly periodic targets 
(TWA\,20) is a confirmed TWA member, the other two having been rejected or needing to  be confirmed.
The rotation period of TWA\,23, although detected in three out  of five time segments, has a critical value
close to the data 1-day observation sampling. Therefore, even if reported in this work, it needs additional observations to be confirmed, therefore, we classify it as \it uncertain.\rm
The data of TWA\,3 were quite sparse
to be suitable for a period search. \\
\indent
To summarise, we have so far derived \it confirmed \rm periods for 15 out of 17 certain late-type members of TW Hya. Rotation periods of TWA\,3A and TWA\,3B are still unknown.\\
\indent
{\bf $\beta$ Pictoris}\\
Data for 13 $\beta$ Pic targets are available in the SuperWASP database, five of  which have been recently identified as $\beta$ Pic members by Kiss et al.\,(\cite{Kiss11}). For seven  targets we confirm the period determined in Paper I. For the other six we determined previously unknown period, five of them classified as \it confirmed, \rm the other one (J01071194-1935359) as \it likely. \rm \\
\indent
To summarise, we derived  \it confirmed \rm periods for 29 of the 37 late-type members of $\beta$ Pic. For a further three stars we  determined \it likely \rm periods, whereas for only one star we have an \it uncertain \rm period. The rotation periods of four stars still remain unknown. \\
\indent
{\bf Tucana/Horologium}\\
Data for 19 Tucana/Horologium  targets are available in the SuperWASP database, two of  which have been recently identified as Tucana/Horologium members by Kiss et al.\,(\cite{Kiss11}).
For nine targets we confirm the rotation periods determined in Paper I. For one target  (J01521830-5950168) we determined the previously unknown period that we classify as \it uncertain.\rm  We revised the period of HIP\,21632, which was taken from the literature in Paper I and based on Hipparcos photometry, and the period of TYC\,8852\,0264\,1 whose earlier determination led to inconsistent $v\sin i$/v$_{\rm eq}$  ratio. This star was excluded in Paper I from rotation period evolution analysis also because it is a rejected member.
SuperWASP data of six targets were quite sparse and unsuited for period search. Data of HD\,25402 were suited but did not allow any
period detection. \\
\indent
To summarise, we derived  \it confirmed \rm periods for 22 of 29 late-type members of Tucana/Horologium. Two stars have rotation periods still  to be confirmed. 
Rotation periods of five stars HIP\,490 HIP\,6856, TYC\,8489\,1155\,1,  AF Hor, and HIP\,16853 are still unknown.\\
\indent
{\bf Columba}\\
Data for 12 Columba  targets are available in the SuperWASP database.
For four targets we confirm the rotation period determined in Paper I. For two targets we determined the previously unknown period that we classify as \it confirmed. \rm We revised
five rotation periods. The revised period of TYC\,7100\,2112\,1 now gives consistent $v\sin i$/v$_{\rm eq}$  ratio. The other four revised periods were in Paper I either  beat periods or detected
in four or fewer  time segments, whereas  they are now well established.  Data of HIP\,25709 were too sparse to allow a period determination.\\
\indent
To summarise, we derived \it confirmed \rm periods for 20 of 23 late-type members of Columba. 
TYC\,6457\,2731\,1 and TYC\,5346\,132\,1 still have  unknown periods. The period of HIP\,25709 is classified as \it likely. \rm\\
\indent
{\bf Carina}\\
We did not find any SuperWASP data for the Carina late-type members.
We derived \it confirmed \rm periods for 14 of 21 late-type members. Four stars have the period classified as \it likely. \rm
The periods of TYC\,8584\,2682\,1, HD107722 is still unknown.  TYC\,8586\,2431\,1 has period leading inconsistent 
$v\sin i$/v$_{\rm eq}$  ratio.\\
\indent
{\bf AB Doradus}\\
Data for 26 AB Doradus  targets are available in the SuperWASP database, and for three targets we determined the previously unknown period (one classified as \it confirmed\rm, and two as \it likely \rm).
For 11 targets we confirm  the rotation period determined in Paper I. We revised six periods, and did not detect any periodicity of five stars although the timeseries are suitable for the period search.  The observations of one target are too sparse  for period search. The rotation period of HIP\,116910 reported in Paper I revealed itself to be the beat period 
of the \it confirmed\rm  P=1.787d discovered in SuperWASP data. The periods of TYC\,7059\,1111\,1 and TYC\,7598\,1488\,1 reported in Paper I led to inconsistent $v\sin i$/v$_{\rm eq}$  ratio, whereas the updated period solved the inconsistency. The period of TYC\,7064\,0839\,1, TYC\,7605\,1429\,1, and TYC\,7627\,2190\,1 reported in Paper I were detected in less than five time segments, whereas the new periods are classified as  \it confirmed\rm. \\
\indent
To summarise, we derived \it confirmed \rm  periods of 29 of 64 late-type members of AB Doradus. 
Twenty  members have periods classified as \it likely\rm. The rotation period of 15 members is still unknown. \\

\indent
In Figs.\,\ref{new_distri_twa}-\ref{new_distri_abdor} we plot the updated rotation period distributions in the young associations studied in Paper I, where the newly discovered and the revised rotation
periods are plotted with squared and squared crossed symbols, respectively, whereas circled symbols represent the rotation periods of the new candidate members
proposed by Kiss et al.\,(\cite{Kiss11}).

Similarly, in Fig.\,\ref{amp_new} we plot the updated distributions of light curve amplitudes versus rotation period in the young associations studied in Paper I.

We notice that both rotation periods and light curve amplitudes of the newly proposed members by Kiss et al.\,(\cite{Kiss11})  agreet with the
values of the other confirmed association members. This evidence gives further support to their assigned membership.

\begin{figure}[t,*h]
\begin{minipage}{11cm}
\includegraphics[trim=0 0 0 0,width=10cm,height=10cm,angle=0]{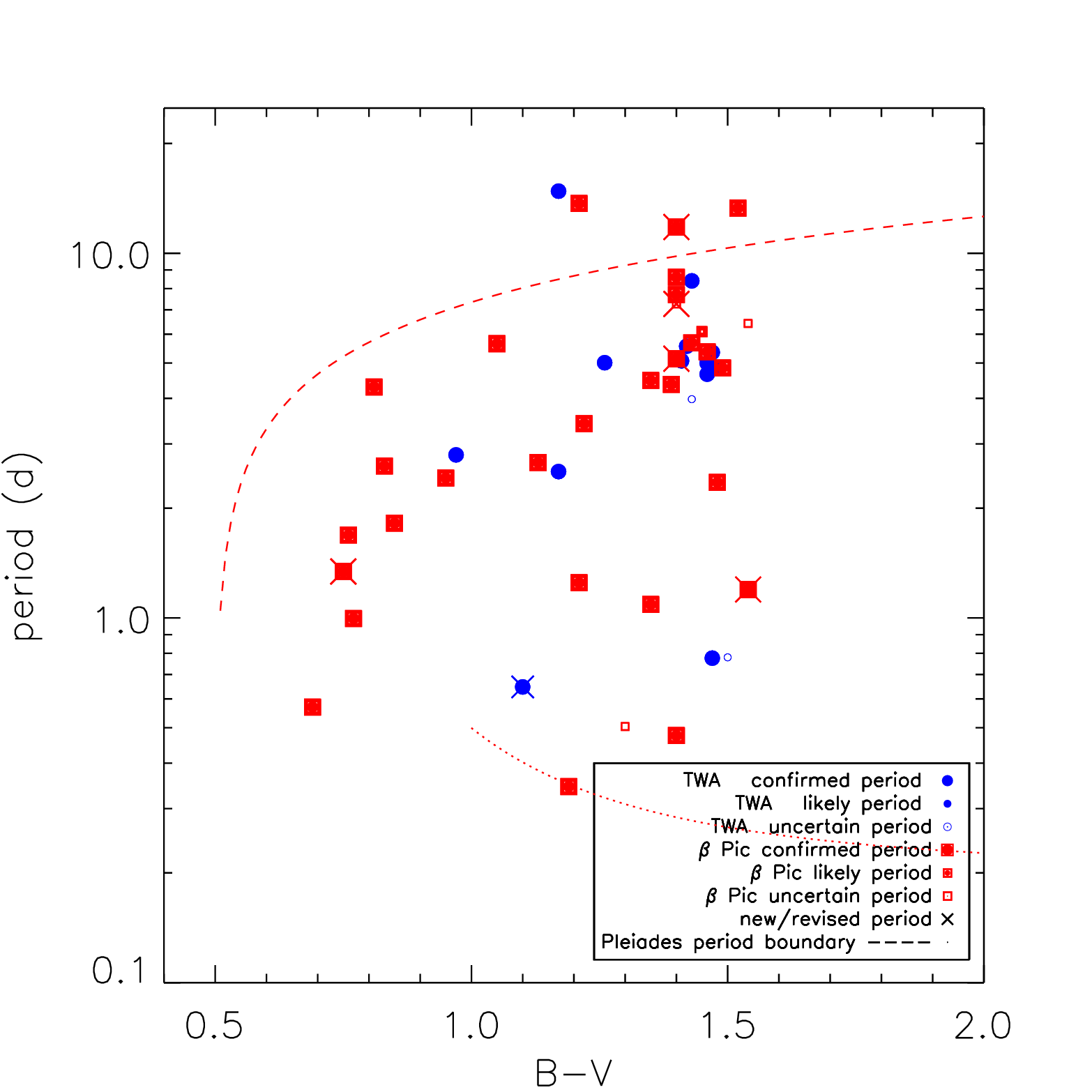}
\end{minipage}
\caption{Updated distributions of rotation periods in TWA + $\beta$ Pic associations The dashed and dotted lines represent  mathematical functions  describing the loci of the upper and lower  bounds of the period distribution of the Pleiades members according to Barnes (\cite{Barnes03}). \label{new_distri_twa}}
\end{figure}

\begin{figure}[t,*h]
\begin{minipage}{11cm}
\includegraphics[trim=0 0 0 0,width=10cm,height=10cm,angle=0]{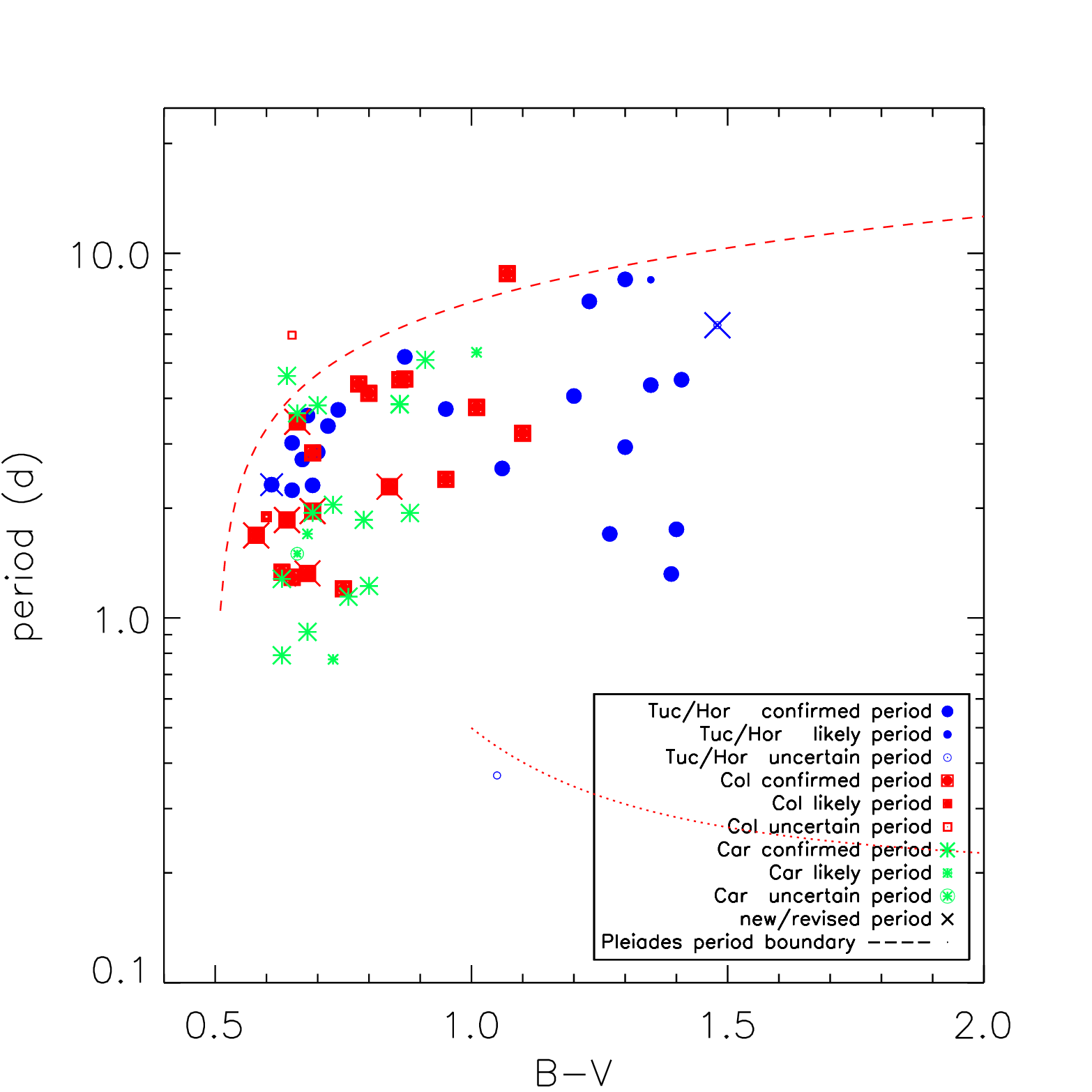}
\end{minipage}
\caption{As in Fig.\,\ref{new_distri_twa} for Tuc/Hor + Car + Col associations. \label{new_distri_tuc}}
\end{figure}

\begin{figure}[t,*h]
\begin{minipage}{11cm}
\includegraphics[trim=0 0 0 0,width=10cm,height=10cm,angle=0]{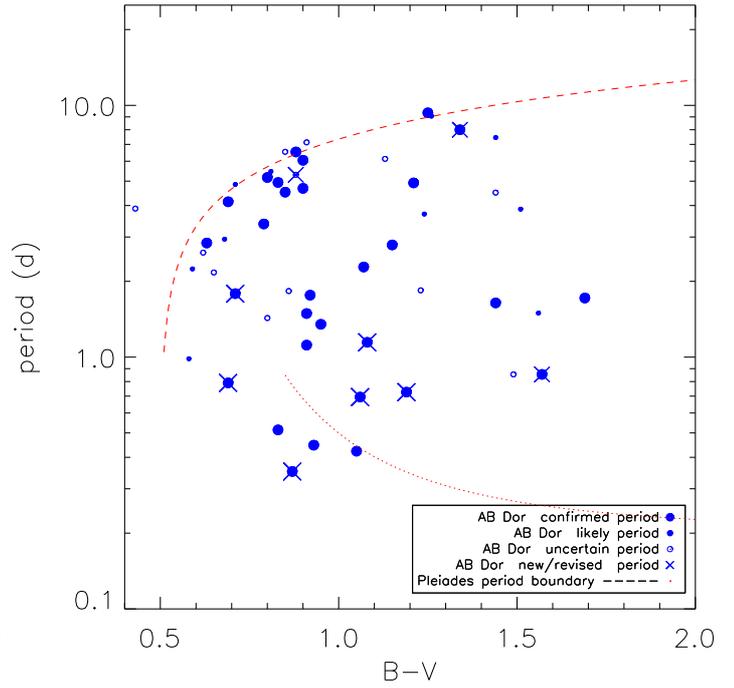}
\end{minipage}
\caption{As in Fig.\,\ref{new_distri_twa} for AB Dor association. \label{new_distri_abdor}}
\end{figure}

\begin{figure}[*t,*h]
\begin{minipage}{11cm}
\includegraphics[trim=0 0 0 0,width=10cm,height=10cm,angle=0]{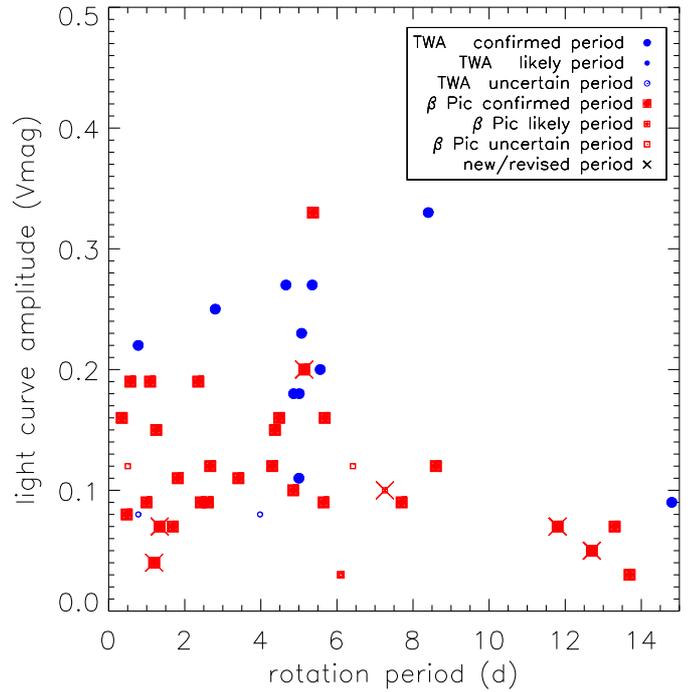}
\end{minipage}
\caption{Distributions of light curve amplitudes in TWA + $\beta$ Pic associations studied in Paper I. \label{amp_distri_twa}}
\end{figure}
\begin{figure}
\begin{minipage}{11cm}
\includegraphics[trim=0 0 0 0,width=10cm,height=10cm,angle=0]{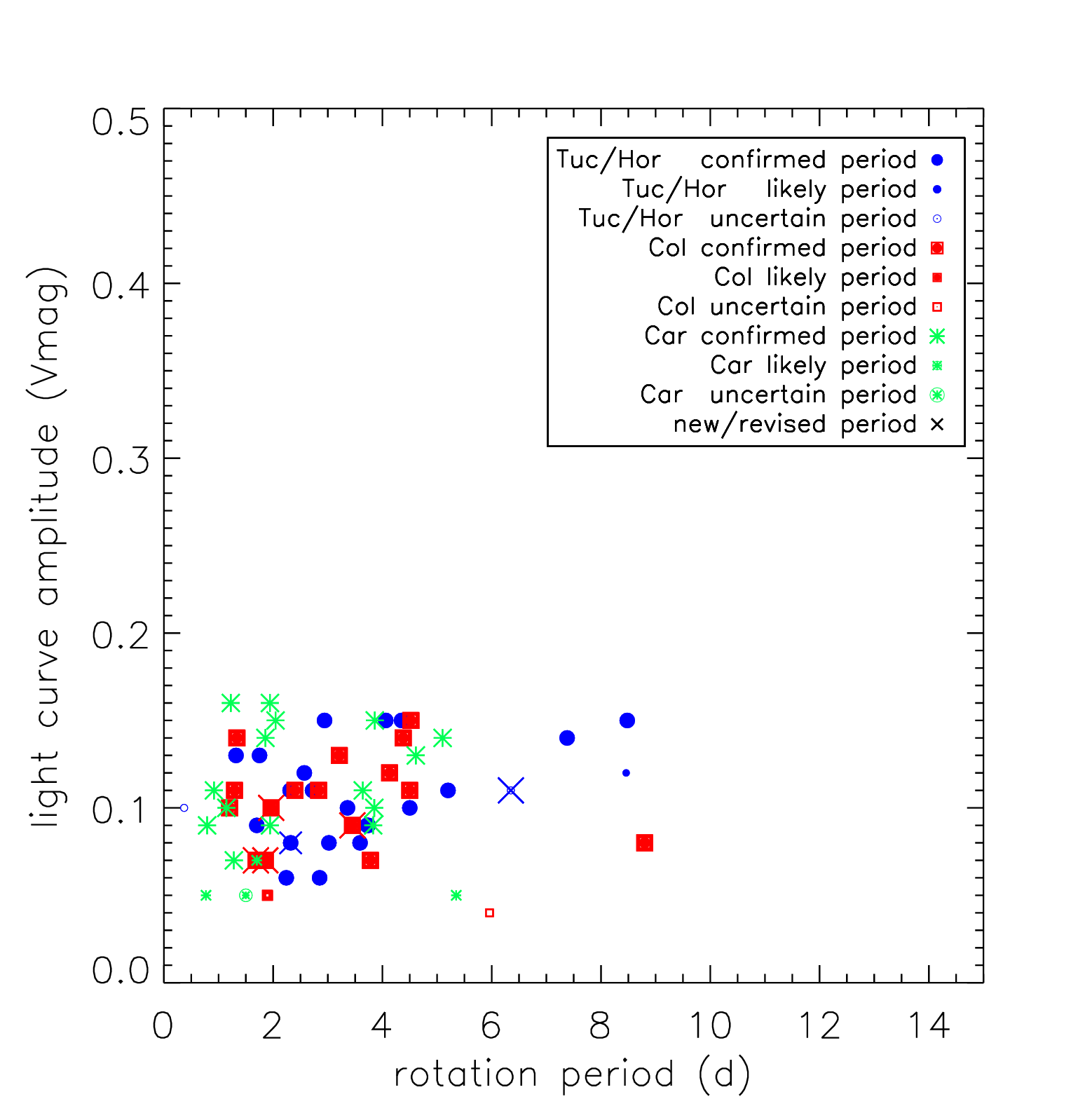}
\end{minipage}
\caption{As in Fig.\,\ref{amp_distri_twa} for Tuc/Hor + Car + Col associations. \label{amp_distri_tuc}}
\end{figure}
\begin{figure}
\begin{minipage}{11cm}
\includegraphics[trim=0 0 0 0,width=10cm,height=10cm,angle=0]{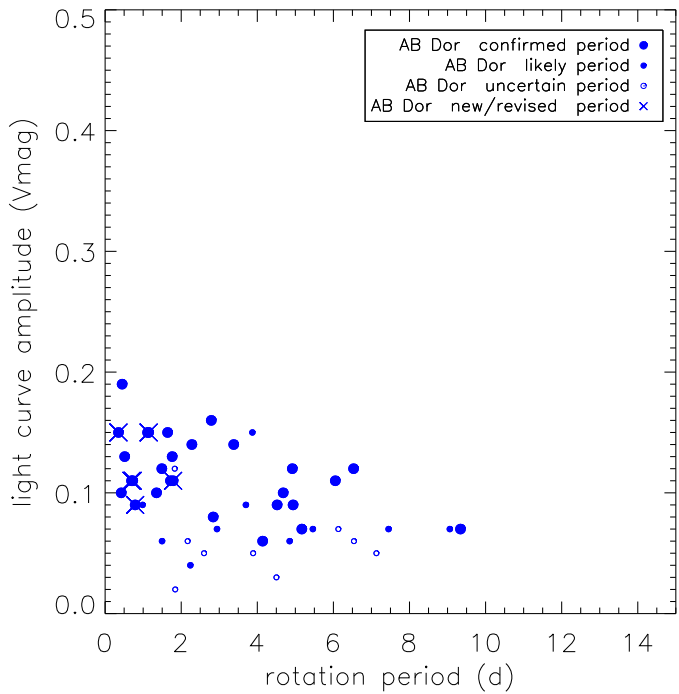}
\end{minipage}
\caption{As in Fig.\,\ref{amp_distri_twa} for AB Dor association. \label{amp_distri_abdor}}
\end{figure}

\begin{table*}
\caption{\label{period-revisited} Summary of period search of 71 targets in Paper I and of seven recently added members with photometry timeseries in the SuperWASP archive.}
\scriptsize
\begin{tabular}{llccll}
\hline
Name & SuperWASP/ASAS ID & P & $\Delta$P & Timeseries	&	Note on Period\\
	&			&	(d)&	(d) & Segments	& 		\\ 
\hline
\multicolumn{6}{c}{\bf TW Hydrae}\\
\hline
\object{TWA\,3AB}		& 1SWASP J111027.90-373152.0	&	...		&	...		& ...		& sparse observations\\
\object{TWA\,6}	& 1SWASP J101828.70-315002.8  	&	0.542	&	0.002	& 4/5	&  confirmed\\
\object{TWA\,8}		& 1SWASP J113241.23-265200.7	&	4.64		&	0.04		& 4/4	&  confirmed\\
\object{TWA\,9}		& 1SWASP J114824.21-372849.2	& 	5.01		&	0.1		& 6/6	&  confirmed\\
\object{TWA\,13}		& 1SWASP J112117.43-344647.1	& 	5.50		&	0.06		& 5/5	&  confirmed\\
\hline

\end{tabular}
\end{table*}

\Online

\begin{deluxetable}{lrcccccl}
\tablecolumns{8}
\tablecaption{\label{tab_period} Summary of period search based on ASAS/superWASP  (SW) photometry. }
\tablewidth{0pt}
 \tablehead{
\colhead {Target}      & \colhead {JD}       & \colhead {JD}    & \colhead {Period} & \colhead {$\Delta$P}  & \colhead {Normalised       } & \colhead {Power at 99\%} & note\\   
\colhead {          }     & \colhead {Initial} & \colhead {Final} & \colhead {(d)}      & \colhead {(d)           }  & {Power} & {Confidence Level}  & \\
}
\startdata
\multicolumn{8}{c}{}\\
\multicolumn{8}{c}{\bf $\epsilon$ / $\eta$ Chamaeleontis}\\
\multicolumn{8}{c}{}\\
\hline
 GSC\,09235-01702  & 1871.8444 & 5039.5551 &  6.86 &  0.09 & 53.57 & 10.10 &   complete \\
 GSC\,09235-01702  & 1874.8509 & 1958.8509 &  6.8 &  0.2 &  9.23 &  6.56 &            \\
 GSC\,09235-01702  & 2623.7775 & 2709.7775 &  6.9 &  0.2 &  9.09 &  4.07 &            \\
 GSC\,09235-01702  & 2709.7775 & 2795.7775 &  6.8 &  0.2 &  7.36 &  4.43 &            \\
 GSC\,09235-01702  & 3034.8547 & 3110.8547 &  6.8 &  0.2 &  9.14 &  5.11 &            \\
\enddata
\end{deluxetable}

\begin{table*}
\scriptsize
\caption{\label{cha_lit} $\epsilon$ Cha association and $\eta$ Cha cluster. Summary data from the literature and mass and radius derived from evolutionary tracks.}
\centering
\begin{tabular}{lr@{\hspace{.1cm}}r@{\hspace{.1cm}}r@{\hspace{.3cm}}r@{\hspace{.1cm}}r@{\hspace{.1cm}}l@{\hspace{.1cm}}r@{\hspace{.1cm}}r@{\hspace{.1cm}}
r@{\hspace{.1cm}}r@{\hspace{.1cm}}r@{\hspace{.1cm}}c@{\hspace{.1cm}}c@{\hspace{.1cm}}c@{\hspace{.1cm}}l@{\hspace{.1cm}}l@{\hspace{.1cm}}l@{\hspace{.1cm}}
}
\hline
  Name        &  \multicolumn{3}{c}{RA } & \multicolumn{3}{c}{DEC} &   \multicolumn{1}{c}{V}      & B$-$V  & V$-$I   &  M$_V$ & \multicolumn{1}{c}{d}      &  \multicolumn{1}{c}{$v\sin i$} & Mass & Radius & Sp.T  &  MGR & n1 \\ 
		&	 \multicolumn{3}{c}{(2000.0)}                    &	 \multicolumn{3}{c}{(2000.0)}                    &  \multicolumn{1}{c}{(mag)}  & (mag)  & (mag) & (mag)   & \multicolumn{1}{c}{(pc)}  &  \multicolumn{1}{c}{(km s$^{-1}$)} & (M$_{\odot}$) & (R$_{\odot}$) & &   &                   \\
\hline

\hline
\multicolumn{18}{c}{}\\
\multicolumn{18}{c}{\bf $\epsilon$ Chamaeleontis}\\
\multicolumn{18}{c}{}\\
\hline

      CP-68\,1388 & 10 & 57 & 49.30 & -69 & 14 &  00.00 & 10.04 &  0.86 &  0.93 &  4.79 & 112 &  25.90 & 1.1 & 1.60 &  K1V(e) & $\epsilon$ Cha & A/0 A$_v$\\
 GSC\,9419-01065 & 11 & 49 & 31.90 & -78 & 51 &  01.00 & 11.90 &  1.27 &  1.80 &  8.31 & 52 &  18.00  & 0.7 & 1.00 &    M0Ve & $\epsilon$ Cha & A/0 A$_v$\\
       HIP\,58285 & 11 & 57 & 13.50 & -79 & 21 & 32.00 & 10.19 &  0.93 &  0.88 &  6.09 & 66 &  39.00  & 0.9 & 0.90 &     K0e & $\epsilon$ Cha & A/2 A$_v$\\
\end{tabular}

\end{table*}

\begin{table*}
\scriptsize
\caption{\label{octans_lit} Octans association. Summary data from the literature and mass and radius derived from evolutionary tracks.}
\centering
\begin{tabular}{lr@{\hspace{.1cm}}r@{\hspace{.1cm}}r@{\hspace{.3cm}}r@{\hspace{.1cm}}r@{\hspace{.1cm}}l@{\hspace{.1cm}}r@{\hspace{.1cm}}r@{\hspace{.1cm}}
r@{\hspace{.1cm}}r@{\hspace{.1cm}}r@{\hspace{.1cm}}c@{\hspace{.1cm}}c@{\hspace{.1cm}}c@{\hspace{.1cm}}l@{\hspace{.1cm}}l@{\hspace{.1cm}}l@{\hspace{.1cm}}
}
\hline
  Name        &  \multicolumn{3}{c}{RA } & \multicolumn{3}{c}{DEC} &   \multicolumn{1}{c}{V$_{\rm min}$}      & B$-$V  & V$-$I   &  M$_V$ & \multicolumn{1}{c}{d}      &  \multicolumn{1}{c}{$v\sin i$} & Mass & Radius & Sp.T  &  MGR & n1 \\ 
		&	 \multicolumn{3}{c}{(2000.0)}                    &	 \multicolumn{3}{c}{(2000.0)}                    &  \multicolumn{1}{c}{(mag)}  & (mag)  & (mag) & (mag)   & \multicolumn{1}{c}{(pc)}  &  \multicolumn{1}{c}{(km s$^{-1}$)} & (M$_{\odot}$) & (R$_{\odot}$) & &   &                   \\
\hline

            CD-58\,860 &  04 & 11 & 55.60 & -58 &  01 & 47.00 &  9.92 &  0.68 &  0.75 &  5.35 &  82 &  20.10 & 1.0 & 1.00 &     G6V & Oct & A/0 \\
          CD-43\,1451 &  04 & 30 & 27.30 & -42 & 48 & 47.00 & 10.75 &  0.79 &  0.90 &  5.35 & 120 &  19.70   & 1.0 & 1.00 &  G9V(e) & Oct & A/0 \\
            CD-72\,248 &  05 &  06 & 50.60 & -72 & 21 & 12.00 & 10.70 &  0.82 &  0.94 &  4.92 & 143 & 190.00 & 1.1 & 1.40 &    K0IV & Oct & A/0 \\
           HD\,274576 &  05 & 28 & 51.40 & -46 & 28 & 18.00 & 10.48 &  0.66 &  0.77 &  5.16 & 116 &  21.10   & 1.0 & 1.00 &     G6V & Oct & A/0 \\
          CD-47\,1999 &  05 & 43 & 32.10 & -47 & 41 & 11.00 & 10.05 &  0.56 &  0.67 &  3.94 & 167 &  39.00   & 1.1 & 1.40 &  G0V(e) & Oct & A/0 \\
  TYC\,7066\,1037\,1 &  05 & 58 & 11.80 & -35 &  00 & 49.00 & 10.77 &  0.80 &  0.84 &  4.90 & 149 &  20.50   & 1.0 & 1.15 &     G9V & Oct & A/1 A$_v$  \\
\end{tabular}

\end{table*}

\begin{table*}
\scriptsize
\caption{\label{argus_lit} Argus association and IC\,2391 cluster. Summary data from the literature and mass and radius derived from evolutionary tracks.}
\centering
\begin{tabular}{lr@{\hspace{.1cm}}r@{\hspace{.1cm}}r@{\hspace{.3cm}}r@{\hspace{.1cm}}r@{\hspace{.1cm}}l@{\hspace{.1cm}}r@{\hspace{.1cm}}r@{\hspace{.1cm}}
r@{\hspace{.1cm}}r@{\hspace{.1cm}}r@{\hspace{.1cm}}c@{\hspace{.1cm}}c@{\hspace{.1cm}}c@{\hspace{.1cm}}l@{\hspace{.1cm}}l@{\hspace{.1cm}}l@{\hspace{.1cm}}
}
\hline
  Name        &  \multicolumn{3}{c}{RA } & \multicolumn{3}{c}{DEC} &   \multicolumn{1}{c}{V$_{\rm min}$}      & B$-$V  & V$-$I   &  M$_V$ & \multicolumn{1}{c}{d}      &  \multicolumn{1}{c}{$v\sin i$} & Mass & Radius & Sp.T  &  MGR & n1 \\ 
		&	 \multicolumn{3}{c}{(2000.0)}                    &	 \multicolumn{3}{c}{(2000.0)}                    &  \multicolumn{1}{c}{(mag)}  & (mag)  & (mag) & (mag)   & \multicolumn{1}{c}{(pc)}  &  \multicolumn{1}{c}{(km s$^{-1}$)} & (M$_{\odot}$) & (R$_{\odot}$) & &   &                   \\
\hline

\hline
\multicolumn{18}{c}{}\\
\multicolumn{18}{c}{\bf Argus}\\
\multicolumn{18}{c}{}\\
\hline
           HD\,5578   & 00 & 56 & 55.50 & -51 & 52 & 32.00 &  9.62 &  0.99 &  1.04 &  6.84 &  36 &  29.3 & 0.8 & 0.90 &  K3Ve & Arg & T/0 \\
          CD-49\,1902 & 05 & 49 & 44.80 & -49 & 18 & 26.00 & 11.30 &  0.69 &  0.78 &  5.68 & 133 &  55.0 & 1.0 & 1.00 &   G7V & Arg & A/0 \\
          CD-56\,1438 & 06 & 11 & 53.00 & -56 & 19 & 05.00 & 11.11 &  0.66 &  0.88 &  6.08 & 101 & 130.0 & 0.9 & 1.00 &   K0V & Arg & T/1 \\
\end{tabular}

\end{table*}

\begin{thebibliography}{}
\bibitem[1995]{Alcala95} Alcal\`a, J. M., Krautter, J., Schmitt, J. H. M. M., et al. 1995, A\&AS, 114, 109
\bibitem[1998]{Baraffe98} Baraffe, I., Chabrier, G., Allard, F., \& Hauschildt, P. 1998, A\&A, 337,403
\bibitem[2009]{Bernhard09} Bernhard, K.,  Bernhard, C.,  \& Bernhard, M. 2009, OEJV, 98, 1
\bibitem[2003]{Barnes03} Barnes, S., 2003, ApJ, 586, 464
\bibitem[2007]{Barnes07} Barnes, S., 2007, ApJ, 669, 1167
\bibitem[2010]{Barnes10} Barnes, S., 2010, ApJ, 722, 222
\bibitem[2007]{Bouvier07} Bouvier, J. 2007 in Star-disk Interaction in Young Stars, Proceedings IAU Symp. 243 eds. J. Bouvier \& I. Appenzeller, p. 231
\bibitem[2008]{Bouvier08} Bouvier, J. 2008, Proceedings of the Annual meeting of the French Society of Astronomy and Astrophysics Eds.: C. Charbonnel, F. Combes and R. Samadi
\bibitem[2010]{Butters10} Butters, O.W., West, R.G., Anderson, D.R., et al. A\&A, 520, L10
\bibitem[2009]{Carpenter09} Carpenter, J.M., Bouwman, J.,  \& Mamajek, E.E, 2009, ApJS, 181, 197
\bibitem[1995a]{Chaboyer95a}Chaboyer, B., Demarque, P., \& Pinsonneault, M. H. 1995a, ApJ, 441, 865
\bibitem[1995b]{Chaboyer95b}Chaboyer, B., Demarque, P., \& Pinsonneault, M. H. 1995a, ApJ, 441, 876
\bibitem[2007]{Chabrier07}Chabrier, G., Gallardo, J., \& Baraffe, I. 2007, A\&A, 472, 17
\bibitem[2010]{Chauvin10} Chauvin, G., Lagrange, A. -M., Bonavita, M., et al. 2010, A\&A, 509, A52
\bibitem[1993]{CollierCameron93}Collier Cameron, A., \& Campbell, C. G. 1993, A\&A, 274, 309
\bibitem[1995]{CollierCameron95}Collier Cameron, A., Campbell, C. G., \& Quaintrell, H. 1995, A\&A, 298, 133
\bibitem[2009]{CollierCameron09}Collier Cameron, A., Davidson, V. A., Hebb, L., et al. 2009, MNRAS, 400, 451
\bibitem[2011]{Desidera11} Desidera, S., Covino, E., Messina, S., et al. 2011, A\&A, 529, A54
\bibitem[2009]{daSilva09} da Silva, L., Torres, C.A.O., de la Reza, R. et al. 2009, A\&A 508, 833
\bibitem[2007]{Doppmann07} Doppmann, G., White, R., Charbonneau, D., \& Torres, G. 2007, Gemini Science 2007, ed. A. Bruch \& I. Fernandez, Poster in Proc. of the 2nd Conf. on Gemini Science Results	
\bibitem[2003]{Feigelson03} Feigelson, E. D., Lawson, W. A., \& Garmire, G. P. 2003, ApJ, 599, 1207
\bibitem[2004]{Feigelson04} Feigelson, E. D., Hornschemeier, A. E., Micela, G., et al. 2004, ApJ, 611, 1107
\bibitem[1998]{Frink98} Frink, S., Roeser, S., Alcal\'a, J.M., et al. 1998, A\&A, 338, 442
\bibitem[2006]{Giampapa06}Giampapa, M. S., Hall, J. C., Radick, R. R., \& Baliunas, S. L. 2006, ApJ, 651, 444
\bibitem[2007]{Guenther07} Guenther, E. W., Esposito, M., Mundt, R. et al. 2007, A\&A, 467, 1147
\bibitem[2010]{James10} James, D. J., Barnes, S. A., Meibom, S., et al. 2010, A\&A 515A, 100 
\bibitem[2005]{Jeffries05}Jeffries, R. D., \& Oliveira, J. M. 2005, MNRAS, 358, 13
\bibitem[2001]{Haisch01}Haisch, K. E., Jr., Lada, E. A., \& Lada, C. J. 2001, ApJ. 553, 153
\bibitem[2010]{Hartman10}Hartman, J. D., Bakos, G. A., Kov\`acs, G., \& Noyes, R. W., 2010, MNRAS, 408, 475
\bibitem[2009]{Hartman09}Hartman, J. D., Gaudi, B. S., Pinsonneault, M. H., et al. 2009, ApJ, 691, 342
\bibitem[1996]{Herbst96} Herbst, W., \& Wittenmyer, R. 1996, BAAS, 28, 1338
\bibitem[2001]{Herbst01}Herbst, W., Bailer-Jones, C. A. L., \& Mundt, R. 2001, ApJ 554, 197
\bibitem[2002]{Herbst02} Herbst, W., Bailer-Jones, C. A. L., Mundt, R., Meisenheimer, K., \& Wackermann, R. 2002, A\&A, 396, 513
\bibitem[2005]{Herbst05} Herbst, W., \& Mundt, R. 2005, ApJ, 633, 967
\bibitem[2007]{Herbst07} Herbst, W., Eisl\"offel, J., Mundt, R., \& Scholz, A. 2007, Protostars and Planets, V. B. Reipurth, D. Jewitt, and K. Keil (eds.), 
University of Arizona Press, Tucson, 951, p.297-311
\bibitem[2006]{Hodgkin06} Hodgkin, S.T., Irwin, J.M., Aigrain, S., et al. 2006, AN, 327, 9 
\bibitem[2009]{Irwin09} Irwin, J.,\& Bouvier, J. 2009, ,The Ages of Stars, Proceedings of the International Astronomical Union, IAU Symposium, Volume 258, p. 363-374
\bibitem[1988]{Kawaler88}Kawaler, S. D. 1988, ApJ, 333, 236
\bibitem[2011]{Kiss11} Kiss, L.L., Moor, A., Szala, T., et al. 2011, MNRAS, 411, 117
\bibitem[1998] {Knude98} Knude, J., \& Hog, E. 1998, A\&A, 341, 451
\bibitem[2001]{Kohler01} Kohler, R. 2001, AJ, 122, 3325
\bibitem[1991]{Koenigl91}Koenigl, A. 1991, ApJ, 370, 39
\bibitem[1997]{Krishnamurthi97}Krishnamurthi, A., Pinsonneault, M. H., Barnes, S., \& Sofia, S. 1997, ApJ, 480, 303
\bibitem[2010]{Lanza10} Lanza, A.F. 2010, A\&A, 512, A77
\bibitem[2004]{Lamm04} Lamm, M. H., Bailer-Jones, C. A. L., Mundt, R., Herbst, W., \& Scholz, A. 2004, A\&A, 417, 557
\bibitem[2001]{Lawson01} Lawson, W.A.,  Crause, L.A., Mamajek, E.E., \& Feigelson, D.E. 2001, MNRAS, 321, 57
\bibitem[2002]{Lawson02} Lawson, W.A.,  Crause, L.A.,  Mamajek, E.E., \& Feigelson, D.E. 2002, MNRAS, 329, 29
\bibitem[2000]{Makarov00}Makarov, V.V., \& Urban, S. 2000, MNRAS, 317, 289
\bibitem[2004]{Makidon04}Makidon, R. B., Rebull, L.M., Strom, S.E., Adams, M.T., \& Patten, B. M. 2004, AJ, 127, 2228
\bibitem[1999]{Mamajek99} Mamajek, E.E., Lawson, W.A., \& Feigelson, E.D. 1999, ApJ, 516, L77
\bibitem[2000]{Mamajek00} Mamajek, E. E., Lawson, W.A., \& Feigelson, E. D. 2000, ApJ, 544, 356
\bibitem[2008]{Mamajek08}Mamajek, E. E., \& Hillenbrand, L. A. 2008, ApJ, 687, 1264
\bibitem[2009]{Marsden09} Marsden, S.C., Carter, B.D., \& Donati, J.-F. 2009, MNRAS 399, 888
\bibitem[1990]{Mathis90} Mathis, J.S. 1990, ARA\&A, 28, 37
\bibitem[1995]{Matt95}Matt, S., \& Pudritz, R. E. 1995, ApJ, 632, 135
\bibitem[2010]{Matt10}Matt, S.P., Pinz\'on, G., de la Reza, R., Greene, T. P. 2010, ApJ, 714, 989
\bibitem[2009]{Meibom09}Meibom, S., Mathieu, R. D., \& Stassun, K.G. 2009, ApJ, 695, 679
\bibitem[2003]{Messina03} Messina, S., Pizzolato, N., Guinan, E. F., \& Rodon\`o, M. 2003, A\&A,410, 671 
\bibitem[2004]{Messina04} Messina, S. Rodon\`o, M., \& Cutispoto, G., 2004, AN, 325, 660 
\bibitem[2007]{Messina07} Messina, S., 2007, Memorie Societ\`a Astron, It., 78, 628
\bibitem[2008]{Messina08} Messina, S., Distefano, E., Parihar, P., et al. 2008, A\&A, 483, 253
\bibitem[2010a]{Messina10} Messina, S., Parihar, P., Koo, J.-R., et al. 2010a, A\&A, 513, A29
\bibitem[2010b]{paper1} Messina, S., Desidera, S., Turatto, M., Lanzafame, A.C., Guinan, E.F. 2010b, A\&A, 520, A15 (Paper I)
\bibitem[1987]{Mestel87}Mestel, L., \& Spruit, H. C. 1987, MNRAS, 226, 57
\bibitem[2009]{Morales09}Morales, J. C.,  Torres, G., Marschall, L. A., \&  Brehm, W. 2009, ApJ, 707, 671
\bibitem[2007]{Norton07} Norton, A. J., Wheatley, P. J., West, R. G., et al. 2007, A\&A, 467, 785
\bibitem[2004]{Pace04}Pace, G., \& Pasquini, L. 2004, A\&A, 426, 1021
\bibitem[1996]{Patten96} Patten, B.M. \& Simon T. 1996, ApJS 106, 489
\bibitem[2007]{Platais07} Platais, I., Melo, C., Mermilliod J.C., et al. 2007, A\&A, 461, 509 
\bibitem[1997]{Pojmanski97} Pojmanski G., 1997, Acta Astronomica, 47, 467
\bibitem[2002]{Pojmanski02} Pojmanski G., 2002, Acta Astronomica, 52, 397
\bibitem[2006]{Pollacco06} Pollacco, D., Skillen, I,, Collier Cameron, A., et al. 2006, PASP 118, 848
\bibitem[2009]{Pont09} Pont, F. 2009, MNRAS, 396, 1789
\bibitem[2009]{Pontoppidan10} Pontoppidan, K. M., Salyk, C., Blake, G,A., et al. ApJ, 2010, 720, 887
\bibitem[1992]{Press92} Press, W. H., Teukolsky, S. A., Vetterling, W. T.,  \& Flannery, B. P. 1992, Numerical recipes in FORTRAN. The art of scientific computing, 
Cambridge: University Press,  2nd ed.
\bibitem[1987]{Radick87}Radick, R. R., Thompson, D. T., Lockwood, G. W., Duncan, D. K., \& Baggett, W. E. 1987, ApJ, 321, 459
\bibitem[2005]{Randich05}Randich, S.,  Bragaglia, A., Pastori, L., et al. 2005, The Messenger, 121, 18
\bibitem[2001]{Rebull01} Rebull, L.M.  2001, AJ, 121, 1676
\bibitem[2002]{Rebull02} Rebull, L. M., Makidon, R. B., Strom, S. E., et al. 2002, AJ, 123, 1528
\bibitem[1982]{Scargle82} Scargle, J.D. 1982, ApJ, 263, 835
\bibitem[2005]{Sestito05}Sestito, P., \& Randich, S. 2005, A\&A, 442, 615
\bibitem[2008]{Setiawan08} Setiawan, J., Weise, P., Henning, Th., et al. 2008, in Precision Spectroscopy in Astrophysics, Edited by N.C. Santos, L. Pasquini, A.C.M. Correia, and M. Romaniello, p. 201 (arXiv 0704.2145)
\bibitem[1994]{Shu94}Shu, F., Najita, J., Ostriker, E., et al. 1994, ApJ, 429, 781
\bibitem[1972]{Skumanich72}Skumanich, A. 1972, ApJ, 171, 565
\bibitem[1999]{Stassun99} Stassun, K.G., Mathieu, R.D., Mazeh, T., \& Vrba, F. 1999, AJ, 117, 2941
\bibitem[2005]{Tamuz05} Tamuz, O., Mazeh, T., \& Zucker, S. 2005, MNRAS, 356, 1466
\bibitem[2003]{Torres03} Torres, C.A.O., Quast, G.R., de la Reza, R., et al. 2003, Astroph. Sp. Sci. Libr. 299, Open Issues in Local Star Formation, 
ed.  J. Lepine \& J. Gregorio-Hetem (Kluwer Academic Publishers), 83
\bibitem[2006]{Torres06} Torres, C.A.O., Quast, G.R., da Silva, L. et al. 2006, A\&A, 460, 695
\bibitem[2008]{Torres08} Torres, C.A.O., Quast, G.R., Melo, C.H.F., \& Sterzik, M.F. 2008, 
  Handbook of Star Forming Regions, Volume II: The Southern Sky ASP Monograph 
  Publications, Vol. 5. Edited by Bo Reipurth, p.757  (arXiv:0808.3362)
\bibitem[1999]{Voges99} Voges, W., Aschenbach, B., Boller, Th. et al. 1999, A\&A, 349, 389
\bibitem[2005]{vonBraun05} von Braun, K., Lee, B.L., Seager, S., et al. 2005, PASP, 117, 141
\bibitem[1967]{Weber67}Weber, E. J., \&  Davis, L. Jr. 1967, ApJ, 148, 217
\bibitem[1985]{Westin85}Westin, T.N.G. 1985, A\&AS, 60, 99
\bibitem[2004]{Zuckerman04} Zuckerman, B. \&  Song, I. 2004, Ann. Rev. Astron. Astr., 42, 685 
\end{thebibliography}
\end{document}